\documentclass[prd,preprintnumbers,superscriptaddress,nofootinbib,showpacs,twocolumn]{revtex4-2}
\usepackage{pstricks}
\usepackage{color}
\usepackage{amssymb,amsmath,bm,bbm}
\usepackage{epsf,epsfig}
\usepackage{afterpage}
\usepackage{longtable}
\usepackage{latexsym,mathrsfs,dsfont}
\usepackage{graphics}
\usepackage{url}
\usepackage{paralist}
\usepackage{bbold}
\usepackage{appendix}

\newcommand{\spur}[1]{\not\! #1 \,}

\newcommand{\np}{\textsc{NP}}
\newcommand{\inter}{\textsc{INT}}

\newcommand{\be}{\begin{equation}}
\newcommand{\ee}{\end{equation}}
\newcommand{\bi}{\begin{itemize}}
\newcommand{\ei}{\end{itemize}}
\newcommand{\ba}{\begin{array}}
\newcommand{\ea}{\end{array}}
\newcommand{\bea}{\begin{eqnarray}}
\newcommand{\eea}{\end{eqnarray}}
\newcommand{\bec}{\begin{center}}
\newcommand{\eec}{\end{center}}
\newcommand{\dd}{\displaystyle}
\newcommand{\nn}{\nonumber}
\newcommand{\qq}{\quad \quad}

\def\@seccntformat#1{\@ifundefined{#1@cntformat}%
   {\csname the#1\endcsname\quad}
   {\csname #1@cntformat\endcsname}
}
\begin{document}

\preprint{BARI-TH/21-726}

\title{  Role of $B_c^+ \to B_{s,d}^{(*)} \, \bar \ell \, \nu_\ell$  in the Standard Model 
and in the search for BSM signals  }
\author{Pietro~Colangelo}
\email[Electronic address:]{pietro.colangelo@ba.infn.it} 
\affiliation{Istituto Nazionale di Fisica Nucleare, Sezione di Bari, via Orabona 4, 70126 Bari, Italy}
\author{Fulvia~De~Fazio}
\email[Electronic address:]{fulvia.defazio@ba.infn.it} 
\affiliation{Istituto Nazionale di Fisica Nucleare, Sezione di Bari, via Orabona 4, 70126 Bari, Italy}
\author{Francesco~Loparco}
\email[Electronic address:]{francesco.loparco1@ba.infn.it} 
\affiliation{Istituto Nazionale di Fisica Nucleare, Sezione di Bari, via Orabona 4, 70126 Bari, Italy}
\affiliation{Dipartimento Interateneo di Fisica "Michelangelo Merlin", Universit\`a degli Studi di Bari, via Orabona 4, 70126 Bari, Italy}

\begin{abstract}
The decays $B_c^+ \to B_{a} \bar \ell  \nu_\ell$ and $B_c^+ \to B_{a}^{*}(\to B_a \gamma) \bar  \ell \nu_\ell$,  with $a=s,d$ and $\ell=e,\mu$,  are studied in the Standard Model (SM) and in the extension  based on the low-energy Hamiltonian comprising the full set of dimension-$6$  semileptonic $c \to s,d$ operators with left-handed neutrinos.  Tests of $\mu/e$  universality are investigated using such modes.
 The heavy quark spin symmetry is applied to relate the relevant hadronic matrix elements and to exploit   lattice QCD results on $B_c$ form factors. 
Optimized observables are selected, and the pattern of  their correlations is studied  to identify  the effects of the various operators in the extended low-energy Hamiltonian.
\end{abstract}


\maketitle

\section{Introduction}
The  $B_c$ meson, first observed by the CDF Collaboration \cite{Abe:1998wi},  is  interesting since it has the  structure of the heavy quarkonium  but it decays weakly. Therefore, this meson is well suited to study both quarkonium and weak interaction features within the same hadronic system. As for weak interactions, in addition to the purely leptonic mode which proceeds through the weak annihilation of the constituent quarks, the  $B_c$ decays  occur through the transitions of  both the charm and beauty quark. The decays induced by the charm transition represent the dominant contribution to the full width despite the smaller available phase-space \cite{Colangelo:1992cx,Beneke:1996xe,Anisimov:1998uk,Kiselev:2000pp}. In our study we focus  on the exclusive semileptonic modes
$B^+_c \to B_{s,d} \bar \ell \nu_\ell$ and $B^+_c \to B^*_{s,d} \bar \ell \nu_\ell$ induced at the quark level by 
$c \to (s,d) \bar \ell \nu_\ell$,  with $\ell=e, \mu$ (the tauonic mode is phase-space forbidden). There are various reasons for such a choice. 

The first one is the possibility of exploiting the heavy quark spin symmetry \cite{Jenkins:1992nb}, which allows us to relate the observables in the modes with final pseudoscalar and vector meson, as well as the different observables in the vector channel. The relatively small phase-space justifies the extrapolation to the full kinematical range of the spin symmetry relations, that strictly hold close to  the zero-recoil point where the produced meson is at rest in the  $B_c$ rest frame \cite{Colangelo:1999zn}. Invoking the heavy quark spin  symmetry the relevant hadronic matrix elements can be expressed in terms of two independent functions, that can be derived from the $B_c \to B_s$ and $B_c \to B_d$ form factors  (FF)    precisely determined by lattice QCD  \cite{Cooper:2020wnj}.

The second reason is the possibility to scrutinize the sensitivity of such processes to  beyond the Standard Model (BSM) effects of the  kind  emerging in $B$ decays, where hints of violation of lepton flavour universality (LFU) are found  \footnote{For recent overviews see \cite{Bifani:2018zmi,Gambino:2020jvv}.}. The measurement of ${\cal B}(B_c \to J/\psi \bar \tau \nu_\tau)$ is also important in this regard \cite{Aaij:2017tyk}.
Such effects can be analyzed in an effective theory framework extending the low-energy SM Hamiltonian that governs the $c \to (s,d) \bar \ell \nu_\ell$ transitions with the inclusion of the full set of semileptonic dimension-$6$ operators with lepton flavour dependent Wilson coefficients.  
The impact of the new operators on the experimental $B_c$  observables can be assessed.  The $D$ and $D_s$ semileptonic decay modes have been recently studied in this context, and the Wilson coefficients of the new operators in the extended Hamiltonian have been constrained using  the available  experimental data \cite{Fajfer:2015ixa,Fleischer:2019wlx,Fuentes-Martin:2020lea,Leng:2020fei,Becirevic:2020rzi}.  The study of 
the sensitivity of this class of $B_c$ decays  to extensions of the Standard Model (the New Physics - NP)  is timely, as  these channels  are accessible at the present facilities.  The  hadronic matrix elements of the new operators can also be given in terms of  the same independent functions entering in the SM ones, invoking the heavy quark spin symmetry. Since the produced $B_s^*$ and $B_d^*$   mesons decay radiatively, we shall provide the expressions of the fully differential $B_c ^+\to B_{s,d}^*(\to B_{s,d}\, \gamma) \bar \ell  \nu_\ell$ decay distribution for the extended low-energy Hamiltonian: such general expressions can also be used for different processes.

 In Sec. \ref{sec:hamil} we introduce the effective  semileptonic Hamiltonian comprising  the full set of dimension-6 operators with left-handed neutrinos,  that generalizes the SM low-energy Hamiltonian. In Sec.  \ref{sec:decay} we provide the decay distributions of $B_c \to B_{s,d} \bar \ell \nu_\ell$ and  $B_c \to B^*_{s,d} (\to B_{s, d} \gamma) \bar \ell \nu_\ell$ obtained from the extended Hamiltonian. In Sec. \ref{HQspin} we discuss the heavy quark spin symmetry  relations  connecting  the SM and NP  operator matrix elements. 
Sec. \ref{sec:num} contains the numerical analysis in SM and a discussion of the effects of the new operators on the $B_c$ decay observables.  The summary and the outlook are presented in the last section.   The  appendices  contain the relations among the hadronic form factors obtained by  the heavy quark spin symmetry (Appendix \ref{app:A}),  and  the coefficient functions of the full angular distribution of the four-body radiative modes $B_c \to B^*_{s,d} (\to B_{s,d} \,\gamma) \bar \ell \nu_\ell$  (Appendix \ref{app:B}). 

\section{ Effective $c \to s,d$ semileptonic  Hamiltonian}\label{sec:hamil}
We consider  the low-energy Hamiltonian  comprising  the full set of   dimension-$6$ semileptonic  $Q \to q$ operators with left-handed neutrinos:
\bea
&&H_{\rm eff}^{Q \to q \bar \ell \nu}= \frac{G_F}{\sqrt{2}} V_{CKM}\nn\\
&&\qq \Big[(1+\epsilon_V^\ell) \left({\bar q} \gamma_\mu  (1-\gamma_5) Q \right)\left(  \bar \nu_{\ell } (1+\gamma_5)   \gamma^\mu \ell \right)\nn \\
&&\qq +  \epsilon_R^\ell \left({\bar q} \gamma_\mu (1+\gamma_5) Q \right)\left( \bar \nu_{\ell} (1+\gamma_5)\gamma^\mu  \ell \right)  \nn \\
&&\qq + \epsilon_S^\ell \, ({\bar q} Q) \left( {\bar \nu_\ell} (1+\gamma_5) \ell \right) \label{hamil}  \\
&&\qq + \epsilon_P^\ell \, \left({\bar q} \gamma_5 Q\right)  \left({\bar \nu_\ell} (1+\gamma_5) \ell \right) \nn \\
&&\qq + \epsilon_T^\ell \, \left({\bar q}  (1+\gamma_5) \sigma_{\mu \nu} Q\right) \,\left( {\bar \nu_\ell}  (1+\gamma_5) \sigma^{\mu \nu} \ell \right)     \Big]  , \nn
\eea
with $Q=c$, and $q$  either  the $s$ or the $d$ quark.   $V_{CKM}$ is the Cabibbo-Kobayashi-Maskawa (CKM) matrix element $V_{cs}$ or $V_{cd}$.
 In addition to  the SM  operator 
 ${\cal O}_{SM}=4 (\bar q_L \gamma^\mu Q_L) \left( {\bar \nu_{\ell L}} \gamma_\mu \ell_L\right)$ 
 and to the operators
  ${\cal O}_S=\left({\bar q} Q \right)\left( {\bar \nu_\ell} (1+\gamma_5) \ell \right)$,
 ${\cal O}_P=\left({\bar q} \gamma_5 Q \right)\left( {\bar \nu_\ell} (1+\gamma_5)  \ell \right)$ and
 ${\cal O}_T=\left({\bar q}  (1+\gamma_5)\sigma_{\mu \nu} Q \right)\left( {\bar \nu_\ell} (1+\gamma_5) \sigma^{\mu \nu}  \ell \right)$,  the operator 
 ${\cal O}_{R}=4 (\bar q_R \gamma^\mu Q_R) \left( {\bar \nu_{\ell L}} \gamma_\mu \ell_L\right)$ 
 is included in Eq.~\eqref{hamil}. It is worth remarking that  in the Standard Model Effective Field Theory the only 
 dimension-$6$ operator with the right-handed quark current is nonlinear in the Higgs field \cite{Buchmuller:1985jz,Cirigliano:2009wk,Aebischer:2020lsx}, and its role  has been the subject of several discussions \cite{Bernard:2006gy,Crivellin:2009sd,Crivellin:2014zpa,Alioli:2017ces,Aebischer:2018iyb,Aebischer:2020lsx}. 
The complex coefficients $\epsilon^\ell_{V,R,S,P,T}$ in the
  low-energy Hamiltonian \eqref{hamil} are  lepton-flavour dependent. 
 
Generalized Hamiltonians as in Eq.~\eqref{hamil} have been studied for $b \to c$ transitions in connection with the anomalies in semileptonic $B \to D^{(*)} \tau \nu_\tau$ decays, obtaining   information on  the various operators \cite{Biancofiore:2013ki,Becirevic:2016hea,Alonso:2016oyd,Colangelo:2016ymy,Jung:2018lfu,Colangelo:2018cnj,Murgui:2019czp,Alguero:2020ukk}. Modes induced by the $b \to u$ induced transition have also been analyzed in such an effective theory approach   \cite{Colangelo:2019axi}.  For both classes  of $b$-quark  transitions,   suitable  observables testing  the Standard Model and challenging LFU  have been identified.   Observables  in baryon decays, in particular in inclusive modes,  have also been studied
 \cite{Colangelo:2020vhu}. Here we focus on the $B_c$  decays governed by the Hamiltonian \eqref{hamil},  to study the SM  phenomenology and to  assess the sensitivity of such channels  to deviations from  the SM.

\section{ Modes $B_c \to P \,\bar \ell  \nu_\ell$ and $B_c \to V(\to P\gamma) \, \bar \ell  \nu_\ell$ }\label{sec:decay}

The  $q^2$  distribution of the $B_c \to P \bar \ell \nu_\ell$  decay, with  $P$ a pseudoscalar meson,  governed by the low-energy Hamiltonian (\ref{hamil})  reads:

\bea
&&
\frac{d \Gamma(B_c \to P  \bar \ell {\nu}_\ell)}{d q^2}=\nn \\
&&\frac{G_F^2 |V_{CKM}|^2 \lambda^{1/2}}{128 \, m_{B_c}^3 \pi^3 q^2 } \left( 1 - \frac{m_\ell^2}{q^2} \right)^2 \nn \\
&&
 \Bigg\{ \left| m_\ell (1 + \epsilon_V^\ell+ \epsilon_R^\ell)  +  \frac{q^2 \epsilon_S^\ell}{m_Q-m_q} \right|^2 (m_{B_c}^2 - m_P^2)^2 f_0^2(q^2)  \nn \\
&&+ \lambda
 \Bigg[ \frac{1}{3} \left| m_\ell (1 + \epsilon_V^\ell+ \epsilon_R^\ell) f_+(q^2) + \frac{4 q^2}{m_{B_c}+m_P} \epsilon_T^\ell f_T(q^2) \right|^2 \nn \\
&&+ \frac{2 q^2}{3} \left| (1 + \epsilon_V^\ell+ \epsilon_R^\ell) f_+(q^2) +4  \frac{m_\ell}{m_{B_c}+m_P} \epsilon_T^\ell f_T(q^2) \right|^2 \Bigg] \Bigg\} . \nn \\   \label{GamBctoP}
\eea
 $G_F$ is the Fermi constant, $q^2$ the squared momentum transferred to the lepton pair and $\lambda=\lambda(m_{B_c}^2,m_P^2,q^2)$ is the triangular function. The form factors $f_{+}$, $f_{0}$  and $f_{T}$ are defined in Appendix \ref{app:A}. The SM expression is recovered setting  to zero all couplings $\epsilon_i^\ell$.

\begin{figure}[b!]
\begin{center}
\includegraphics[width = 0.55\textwidth]{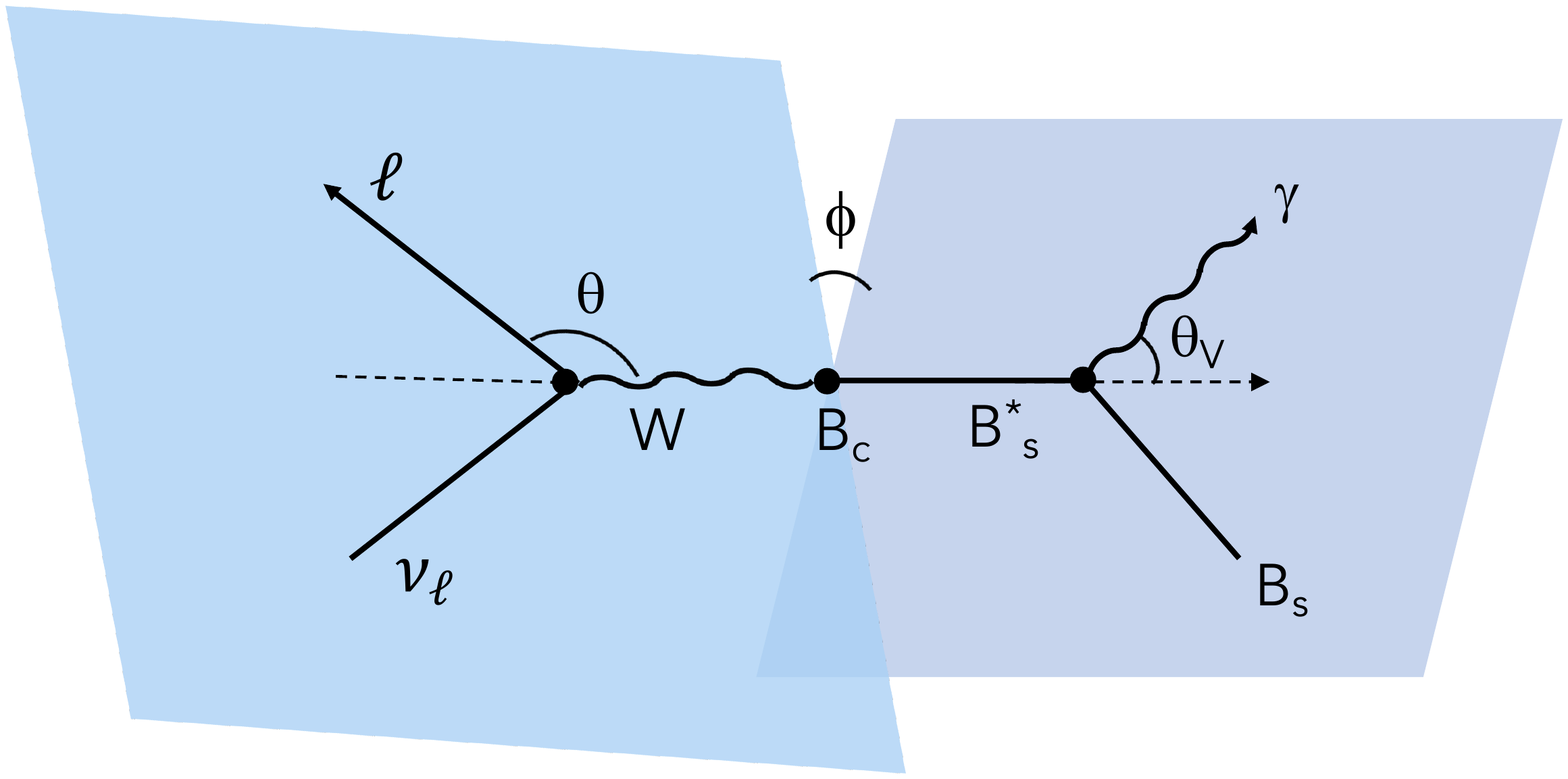} 
\vspace*{-1cm}
\caption{ \baselineskip 10pt  \small  Kinematics of the $B_c \to B_s^* (B_s \gamma) \bar \ell \nu_\ell$ decay. }\label{fig:piani}
\end{center}
\end{figure}

In the case of a final vector meson $V$ decaying to $P \gamma$, namely  $B^*_{s,d}$,  the four-body kinematics of   $B_c \to V(\to P\gamma) \,\bar \ell  \nu_\ell$ is shown in Fig.~\ref{fig:piani}. 
The fully differential decay width is expressed in terms of  $q^2$ and  of the angles $\theta_V$, $\theta$ and $\phi$  defined  in the figure:
\bea
&&\frac{d^4 \Gamma ( B_c \to V(\to P\gamma) \bar \ell  \nu_\ell)}{dq^2  \,d\cos \theta_V \,d\cos \theta \,d\phi}  ={\cal N_\gamma}|{\vec p}_{V}| \left(1- \frac{ m_\ell^2}{q^2}\right)^2 \nn \\
&&\qq \Big\{I_{1s} \,\sin^2 \theta_V+I_{1c} \,(3+\cos 2\theta_V )\nn \\ 
&&\qq + (I_{2s} \,\sin^2 \theta_V+I_{2c} \,(3+\cos 2\theta_V )) \cos 2\theta \nn \\   
&&\qq +I_3 \,\sin^2 \theta_V \sin^2 \theta  \cos  2 \phi +I_4 \sin 2 \theta_V \sin 2\theta \cos  \phi  \nn \\ 
&&\qq +I_5 \, \sin 2 \theta_V  \sin \theta \cos  \phi   \label{angulargamma}\\
&&\qq +(I_{6s} \, \sin^2 \theta_V+I_{6c} \,(3+\cos 2\theta_V ))\cos \theta
 \nn \\
&&\qq + I_7 \sin 2 \theta_V \sin \theta \sin  \phi+I_8\sin 2 \theta_V \sin 2\theta \sin  \phi \nn \\
&&\qq +I_9   \,\sin^2 \theta_V \sin^2 \theta  \sin  2 \phi \Big\}, \nn
\eea
with $|{\vec p}_{V}|=\sqrt{\lambda(m_{B_c}^2,m_V^2,q^2)}/2 m_{B_c}$.
The distribution \eqref{angulargamma} is obtained in the narrow width approximation for the meson $V$, and the factor    
${\cal N}_\gamma=\displaystyle{\frac{3G_F^2 |V_{CKM}|^2 {\cal B}(V \to P \gamma)}{128(2\pi)^4m_{B_c}^2}}$
comprises the   $V \to P \gamma$ branching fraction.
The angular coefficient functions $I_i(q^2)$ encode  the   dynamics and the SM and of   NP  
described by the  Hamiltonian \eqref{hamil}.
We provide them  for the full set of operators, generalizing the results obtained in     \cite{Colangelo:2018cnj} for the tensor operator:

\bea
I_i &=& |1+\epsilon_V|^2 \,I_i^{SM}+|\epsilon_R|^2I_i^{NP,R}+|\epsilon_P|^2I_i^{NP,P} \hspace*{1cm} \nn \\
&+&|\epsilon_T|^2I_i^{NP,T} +2 \, {\rm Re}\left[\epsilon_R(1+\epsilon_V^* )\right] I_i^{INT,R} \nn \\
&+&2 \, {\rm Re}\left[\epsilon_P(1+\epsilon_V^* )\right] I_i^{INT,P} \nn \\
&+&2 \, {\rm Re}\left[\epsilon_T(1+\epsilon_V^* )\right] I_i^{INT,T} \label{eq:Iang1} \\
&+&2 \, {\rm Re}\left[\epsilon_R \epsilon_T^* \right] I_i^{INT,RT} 
+2 \, {\rm Re}\left[\epsilon_P \epsilon_T^* \right] I_i^{INT,PT}\nn \\
&+&2 \, {\rm Re}\left[\epsilon_P \epsilon_R^* \right] I_i^{INT,PR} \nn
\eea
for $ i=1,\dots 6$,
\bea
I_7 &=&2 \, {\rm Im}\left[\epsilon_R(1+\epsilon_V^* )\right] I_7^{INT,R} \nn \\
&+&2 \, {\rm Im}\left[\epsilon_P(1+\epsilon_V^* )\right] I_7^{INT,P}\nn \\
&+&2 \, {\rm Im}\left[\epsilon_T(1+\epsilon_V^* )\right] I_7^{INT,T} \label{eq:Iang2} \\
&+&2 \, {\rm Im}\left[\epsilon_R \epsilon_T^* \right] I_7^{INT,RT}+2 \, {\rm Im}\left[\epsilon_P \epsilon_T^* \right] I_7^{INT,PT} \qq \nn \\
&+&2 \, {\rm Im}\left[\epsilon_P \epsilon_R^* \right] I_7^{INT,PR}  , \nn
\eea
and
\be
I_i=2 \, {\rm Im}\left[\epsilon_R (1+\epsilon_V^* ) \right] I_i^{INT,R} 
\label{eq:Iang3} \ee
for $ i=8,\,9$.
In SM the angular coefficient functions are given in terms of the  helicity amplitudes

\bea
H_0 &=&\frac{1}{{2m_V(m_{B_c}+m_V) \sqrt{q^2}}} \nn \\
&& \Big( (m_{B_c}+m_V)^2(m_{B_c}^2-m_V^2-q^2) A_1(q^2)\nn \\
&-&\lambda(m_{B_c}^2,\,m_V^2,\,q^2) A_2(q^2) \Big) \label{HampV}  \\
H_\pm&=& \frac{(m_{B_c}+m_V)^2 A_1(q^2)\mp\sqrt{\lambda(m_{B_c}^2,\,m_V^2,\,q^2)}V(q^2)}{m_{B_c}+m_V}  \nn  \\
H_t&=& -\frac{\sqrt{\lambda(m_{B_c}^2,\,m_V^2,\,q^2)}}{\sqrt{q^2}} \,A_0(q^2) . \,\,\,  \nn
\eea
For the NP operators  the following amplitudes are also introduced:
\bea
H_\pm^{NP} &=&
 \frac{1}{\sqrt{q^2}}\Big\{\Big(m_{B_c}^2-m_V^2 \pm \sqrt{\lambda(m_{B_c}^2,m_V^2,q^2)} \Big)(T_1+ T_2) \nn \\
&+&q^2(T_1- T_2)\Big\} \nn \\
H_L^{NP}&=& 4\Big\{
\frac{\lambda (m_{B_c}^2,m_V^2,q^2)}{m_V(m_{B_c}+m_V)^2} \, T_0+2\frac{m_{B_c}^2+m_V^2-q^2}{m_V} T_1\nn \\
&+&4 m_V T_2\Big\}  . \label{HampNP}
\eea
The form factors $V$, $A_i$ and $T_i$ are defined  in Appendix \ref{app:A}. The coefficient functions  in Eqs.~\eqref{eq:Iang1},  \eqref{eq:Iang2} and  \eqref{eq:Iang3}, expressed in terms of the amplitudes  \eqref{HampV} and \eqref{HampNP}, are collected in  Appendix \ref{app:B}. With such expressions the various observables can be computed by suitable integrations of the distribution in Eq.~\eqref{angulargamma}.

\section{Heavy quark spin symmetry and relations among form factors }\label{HQspin}
In the infinite heavy quark mass limit  $m_Q \gg \Lambda_{\rm QCD}$ the QCD Lagrangian exhibits a heavy quark (HQ) spin symmetry,  with the decoupling of the heavy quark spin from gluons  \cite{Neubert:1993mb}. This produces the decoupling of  the spins of the heavy quarks in $B_c$: the spin-spin interaction vanishes in this limit. Important consequences of the HQ spin symmetry are the relations among the  form factors parametrizing  the weak current matrix elements  of $B_c$ and mesons comprising a single heavy quark ($B_s^{(*)},  B_d^{(*)},  D^{(*)},\dots$) or two heavy quarks  ($\eta_c, J/\psi, \psi(2S), \dots$)    
  \cite{Jenkins:1992nb}.
 
In the semileptonic $B_c \to B_a^{(*)}$   ($a=s,d$)  decays induced by the $c \to s,d$ transition, since $m_c \ll m_b$ the energy released to the final hadronic system is much smaller than $m_b$.   The $b$ quark remains almost unaffected, so that the final meson  keeps the same $B_c$ four-velocity $v$. 
Denoting the initial and final meson four-momenta  as
$p=m_{B_c} v$ and $p^\prime=m_{B_a} v^\prime =m_{B_a} v +k$  ,  with $k$   a small residual momentum, the four-momentum transferred to the leptons is $q=p-p^\prime=(m_{B_c}-m_{B_a})v-k$, with  $v \cdot k={\cal O}(1/m_b)$.

The relations stemming from the  HQ spin symmetry can be worked out using the trace formalism \cite{Falk:1990yz}. The heavy pseudoscalar and vector mesons are collected in doublets, the two components of which represent states  differing only for the orientation of the heavy quark spins.
The  $B_c^+$ and $B_c^{*+}$ doublet   comprising the heavy $c$ and $ \bar b$ quarks  is described by the effective fields
\be
H^{c \bar b}= \frac{1+\spur{v}}{2} \left[B_c^{*\mu} \gamma_\mu - B_c \gamma_5 \right] \frac{1-\spur{v}}{2}  .\label{bc-doublet}
\ee
 The   $B_a$ and $B_a^*$ doublet ($a$ an $SU(3)_F$ index) with the single heavy antiquark $\bar b$ is described by  the effective fields
\be
H^{\bar b}= \left[B_a^{*\mu} \gamma_\mu - B_a \gamma_5 \right] \frac{1-\spur{v}}{2} . \label{qb-doublet}
\ee
$B$ and $B^*$ are operators that  include a factor $\sqrt{m_B}$ and $\sqrt{m_B^*}$ and have   dimension $3/2$. The equations 
${\spur v}  H^{c \bar b}=H^{c \bar b}$, $\ H^{c \bar b} {\spur v}=-H^{c \bar b}$, ${\spur v}  H^{\bar b}=H^{\bar b}$,  $\ H^{ \bar b} {\spur v}=-H^{ \bar b}$ are satisfied.
Under the heavy quark spin transformations and  light quark $SU(3)_F$ transformations  the doublets transform as 
\bea
H^{c \bar b} \to S_c H^{c \bar b} S^\dagger_b \nn \\
H^{\bar b}_a \to (U H^{\bar b})_a S^\dagger_b . \label{eq:transf}
\eea
 The  matrix elements  of the quark current ${\bar q} \Gamma Q$ between 
$B_c$ and $B_a^{(*)}$, with  $\Gamma$ a generic  product of Dirac matrices,  can be written  as
\bea
&&
\langle B_a^{(*)}(v,k)| {\bar q} \Gamma Q| { B_c}(v) \rangle =\nn \\
&&
-\sqrt{ m_{B_c} m_{B_a}} \,  {\rm Tr} \left[ \overline H_a^{(\bar b)} \Omega_a(v, a_0 k) \Gamma H^{(c \bar b)}\right] , \label{omega}
\eea
with $\overline H_a=\gamma^0 H_a^\dagger \gamma^0$ and   
 are invariant under rotations of the $\bar b$  spin.  The most general matrix depending on $v$ and $k$ is
\be
\Omega_a(v, a_0 k)=\Omega_{1a}+ \spur{k} a_0\Omega_{2a} . \label{omega1}
\ee
 It involves two dimensionless nonperturbative  functions, the form factors $\Omega_{1a}$ and $\Omega_{2a}$. The dimensionful parameter $a_0$ can be identified with the  length scale of the process,  typically the Bohr radius of the mesons.  
At odds with the weak matrix elements of mesons comprising a single heavy quark,  that are expressed in terms of  a single universal function (the Isgur-Wise function \cite{Isgur:1989vq,Isgur:1989ed}) normalized to 1 at the zero-recoil point $v \cdot v^\prime =1$ due to  the heavy quark flavour symmetry,  no normalization is fixed for $\Omega_{1}$ and $\Omega_{2}$. Such form factors encode the QCD dynamics and  must be determined by  nonperturbative  methods. 

The SM matrix elements relevant for  $B_c^+ \to B_a \ell^+ {\nu}_\ell$ involve the form factors   $f_+^{B_c \to P_a}$ and $f_0^{B_c \to P_a}$ defined in (\ref{BctoP}). On the other hand, four form factors are needed in SM for each  $B_c \to B_a^* \ell {\bar \nu}_\ell$ mode,  $V^{B_c \to V_a}$ and $A_{1,2,0}^{B_c \to V_a}$ defined in (\ref{BctoV}). They  parametrize the 
hadronic  matrix elements of the SM operator in the low-energy Hamiltonian (\ref{hamil}). The  matrix elements of the operators  with a scalar and pseudoscalar quark current in Eq.~(\ref{hamil}) do not involve  new form factors:  the scalar operator contributes only to $B_c \to B_a \ell {\bar \nu}_\ell$ and its hadronic  matrix element is given in terms of   $f_0^{B_c \to B_a}$ and of the masses of the  quarks involved in the transitions.  The pseudoscalar operator contributes only to  $B_c \to B_a^* \ell {\bar \nu}_\ell$ and its matrix element can be expressed in terms of $A_0^{B_c \to B_a^*}$ and the  quark masses  (Appendix \ref{app:A}). The matrix elements of the tensor operator in (\ref{hamil})  require  the form factors $f_T^{B_c \to B_a}$  for $B_c^+ \to B_a \ell^+ {\nu}_\ell$ and  $T_{1,2,0}^{B_c \to B_a^*}$ for  $B_c^+ \to B_a^* \ell^+ {\nu}_\ell$ defined in  Appendix \ref{app:A}.

Exploiting the HQ spin symmetry  all the form factors $f_+,\,f_0,\,f_T$ and $V,A_i,T_i$ can be given  in terms of the  functions $\Omega_{1,2}$ in \eqref{omega1}.
Such relations can be inverted to express $\Omega_1$ and $\Omega_2$ in terms of $f_+$ and $f_0$,
Eq.~\eqref{eq:om12},  and can be used once such functions are determined in a nonperturbative way. All relations are  in Appendix \ref{app:A}.  
 The result is that  $f_+$ and $f_0$, accompanied with the relations from the HQ spin symmetry,    provide enough information to study  the full phenomenology of the $B_c \to B^{(*)}_a$ semileptonic modes in SM and beyond.  
  
The relations  among the form factors are valid close to the zero-recoil point, at maximum momentum  squared transferred to the lepton pair $q^2_{max}=(m_{B_c}-m_{B_a^{(*)}})^2$. However,  since the phase space  for  $B_c \to B_a^{(*)}$  is  small,   such relations  can be extrapolated  to the full kinematical $q^2$ range. The assumption  can be  checked  once other form factors are available, by a comparison with the  expressions in the heavy quark limit.

\section{Numerical analysis}\label{sec:num}
We  describe  several observables in   $B_c^+ \to B_{s,d}  \ell^+ \nu_\ell$ and $B_c^+ \to B_{s,d}^*(\to B_{s,d} \gamma) \ell^+ \nu_\ell$ in the Standard  Model.  We also study their sensitivity 
 to the BSM operators in the low-energy Hamiltonian. 

For the hadronic matrix elements of the various operators   in Eq.~\eqref{hamil} we  exploit  the HQ spin symmetry and express  all form factors in terms of the universal  functions $\Omega_{1s(d)}$ and  
$\Omega_{2s(d)}$ using the relations in Appendix \ref{app:A}.   $\Omega_{1s(d)}$ and  
$\Omega_{2s(d)}$ are determined  from the form factors 
$f_{+,0}^{B_c \to B_{s}}$ and $f_{+,0}^{B_c \to B_{d}}$ computed by lattice QCD in Ref.~\cite{Cooper:2020wnj}. 
In such computation the form factors are evaluated in the full $q^2$ range, by a chain fit of the results obtained by a non-relativistic QCD  treatment of the $b$ quark and by using the highly improved staggered quark method. The variable $t=q^2$, with kinematical bound  $m_\ell^2 \le t \leq t_-=(m_{B_c}-m_{B_{s(d)}})^2$, is mapped into the variable
$z(t)=\frac{\sqrt{t_+-t}-\sqrt t_+}{\sqrt{t_+-t}+\sqrt t_+}$ with $t_+=(m_{B_c}+m_{B_{s(d)}})^2$  chosen to be   larger than the lowest threshold for hadron production in the $t$ channel, the  $D K$ and $D \pi$ threshold. To optimize the 
calculation, a rescaled variable 
 $z_p(t)=z(t)/z(M^2_{res})$ is defined, with $M_{res}$ a  suitably chosen mass parameter. Each form factor $f(t)$  is  expressed (in the continuum limit of the lattice discretization) as a truncated  power series of $z_p$:
\be
f(t)=P(t) \sum_n^N  A_n z_p(t)^n \,\, ,
\ee
with $P(t)$ a  function chosen to describe the main computed $t$-dependence. As a result,  each form factor is  determined by the set of coefficients $A_n$ together with  their errors and  error correlation matrices.  The  functions $\Omega_1(y)$ and $a_0 \Omega_2(y)$ obtained  for  the   $c \to s$ and $c \to d$ transitions are depicted in Fig.~\ref{fig:omega} together with their uncertainties. They are expressed  in terms of   the variable $y=\displaystyle\frac{p \cdot p^\prime}{m_{B_c}m_{B_a}}=\frac{m_{B_c}^2+m_{B_a}^2-q^2}{2 m_{B_c}m_{B_a}}$ in the range $[1,y_{max}]$, with $y_{max}$ corresponding to $q^2_{min}=m_\ell^2$. The numerical  values of the other parameters, taken from  the Particle Data Group \cite{Zyla:2020zbs},  are listed in Table \ref{tab:par}.

\begin{widetext}
\begin{center}
\begin{figure}[t!]
\begin{center}
\includegraphics[width = 0.8\textwidth]{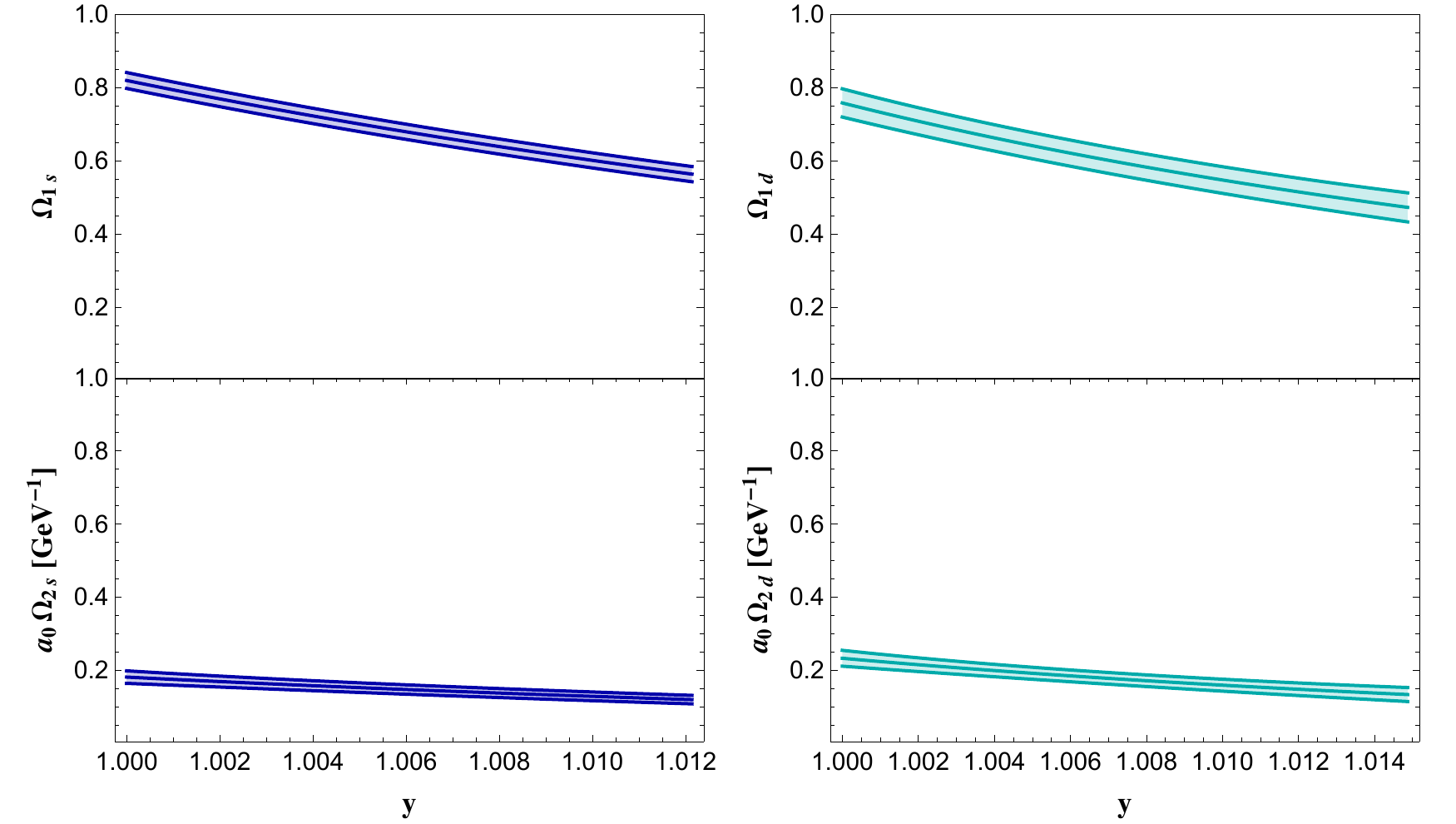} 
\caption{ \baselineskip 10pt  \small   Universal  functions $\Omega_{1}(y)$ (top) and $a_0 \Omega_{2}(y)$   (bottom panels) obtained using Eq.~\eqref{eq:om12}  and the form factors $f_+$ and $f_0$ computed in  Ref.~\cite{Cooper:2020wnj} for  $B_c \to B_s$  (left) and  $B_c \to B_d$   matrix elements (right panels),   with  $\dd y=\frac{p \cdot p^\prime}{m_{B_c}m_{B_a}}$. }\label{fig:omega}
\end{center}
\end{figure}
\end{center}
\end{widetext}
%

\begin{table}[t]
\caption{ \small  Parameters,  from Ref.~\cite{Zyla:2020zbs}.}\label{tab:par}
\vspace{0.3cm}
\centering
\begin{tabular}{cc}
\hline
$m_{B_c}$ &  $6274.9\pm0.8$ MeV \\
$\tau_{B_c}$  &   $(0.510\pm0.009) \times 10^{-12}$ s\\
$m_{B_s}$  & $5366.88\pm0.14$ MeV \\
$m_{B_s^*}$  & $5415.8\pm1.5$ MeV \\
${\cal B}(B_s^*\to B_s \gamma)$  &1  \\
$m_{B_d}$  & $5279.63\pm0.20$ MeV\\
$m_{B_d^*}$  &$5324.7\pm0.21$ MeV \\
${\cal B}(B_d^*\to B_d \gamma)$  &1  \\
$|V_{cs}|$  & $0.987\pm0.011$ \\
$|V_{cd}|$  & $0.221\pm0.004$ \\
$m_d^{\overline{MS}}(2 \, GeV)$ & $4.67^{+0.48}_{-0.17}$ MeV \\
$m_s^{\overline{MS}}(2 \, GeV)$ & $93^{+11}_{-5}$ MeV \\
$m_c$ & $1.67\pm0.07$ GeV\\
\hline
\end{tabular}
\end{table}

The analysis of the sensitivity to the BSM operators  in Eq.~(\ref{hamil}) requires a set of input values for the coefficients $\epsilon_i^\ell$. There are experimental constraints,  in particular  from the purely leptonic $D_s$  and $D^+$   decay widths, from  the  semileptonic $D^{0(+)}$ decays to $K^{-(0)},K^{*-(0)}$ and $\pi^{-(0)},\rho^{-(0)}$, and from  the  semileptonic $D_s \to \phi$ transitions \cite{Fajfer:2015ixa,Fleischer:2019wlx,Leng:2020fei,Becirevic:2020rzi}. Ranges of values have been determined upon the assumption that all  $\epsilon_i^\ell$ are real  \cite{Becirevic:2020rzi}:    $\epsilon_V^\mu=(1.65 \pm 2.02) \times 10^{-2} $,  $\epsilon_R^\mu=(-1.35 \pm 2.02) \times 10^{-2}$,  $\epsilon_S^\mu =(-1.0\pm 2.0) \times 10^{-2} $,  $\epsilon_P^\mu=(0.9 \pm 1.4) \times 10^{-3}$  and $\epsilon_T^\mu=(1.2 \pm 1.8) \times 10^{-2}$
for the $c \to s$ transition, and 
  $\epsilon_V^\mu=(5.0 \pm 2.1) \times  10^{-2} $, $\epsilon_R^\mu=(2.0 \pm 2.0) \times 10^{-2}$, $\epsilon_S^\mu=(-9.0\pm 7.0) \times 10^{-2} $, $\epsilon_P^\mu=(-2.6 \pm 1.3) \times 10^{-3}$  and $\epsilon_T^\mu=(-2.0 \pm 1.4)\times 10^{-1}$  for  the $c \to d$ transition.   Interestingly, the allowed range for $\epsilon_T^\mu$ in the $c \to d$ transition is wide.  We  vary the couplings in these intervals with the purpose of describing the effects of  the various NP operators.
Assuming a hierarchy in LFU violation, all  couplings  for the electron operators
 $\epsilon^e_{V,R,S,P,T}$ are kept to zero, hence such modes are only described in SM.

\subsection{$B_c \to B_s \ell^- \bar \nu_\ell$ and $B_c \to B_s^*(\to B_s \gamma) \ell^- \bar \nu_\ell$ }
The semileptonic $B_c$ decays  induced  by the $c \to s$ transition are expected to constitute the largest fraction of semileptonic modes 
\cite{Colangelo:1999zn,Ivanov:2000aj,Ebert:2003wc,Kiselev:2003mp,Ivanov:2006ni,Hernandez:2006gt,Wang:2008xt,Choi:2009ai,Barik:2009zz,Dhir:2009ub,Chang:2014jca,Shi:2016gqt}.  
The prediction in SM 
\be
{\cal B}(B_c^+ \to B_s \,\mu^+  \nu_\mu)=0.0125\,(4)\, \left(\frac{|V_{cs}|}{0.987}\right)^2    \,\,\label{BrBsRis}
\ee
follows from the use of form factors  in \cite{Cooper:2020wnj}. The quoted error refers only to the form factor uncertainties, the errors from  the CKM matrix element and from the $B_c$ lifetime in Table \ref{tab:par} can  be simply  added, the error from the mass parameters is small. For the electron mode  the result is: 
\be
{\cal B}(B_c^+ \to B_s \, e^+  \nu_e)=0.0131\,(4)\, \left(\frac{|V_{cs}|}{0.987}\right)^2    . \label{BrBsRise}
\ee

\begin{figure}[t]
\begin{center}
\includegraphics[width = 0.45\textwidth]{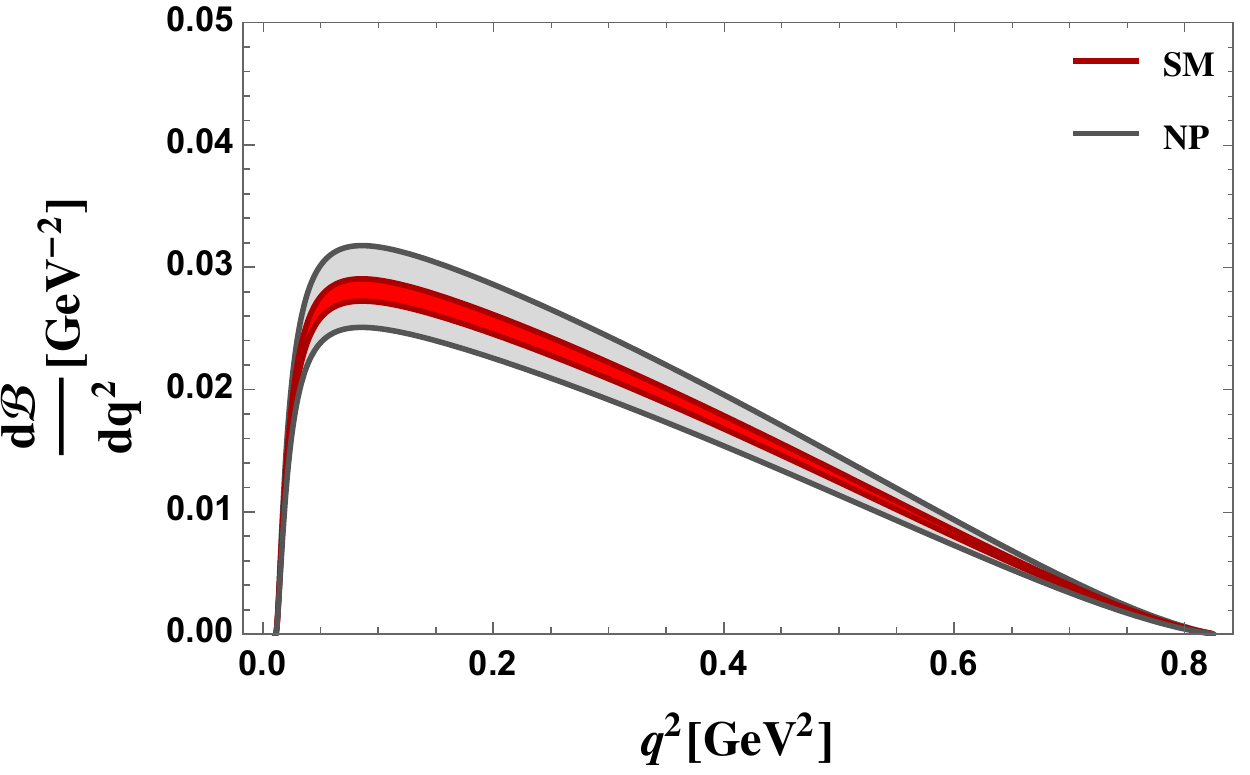} \\
\includegraphics[width = 0.45\textwidth]{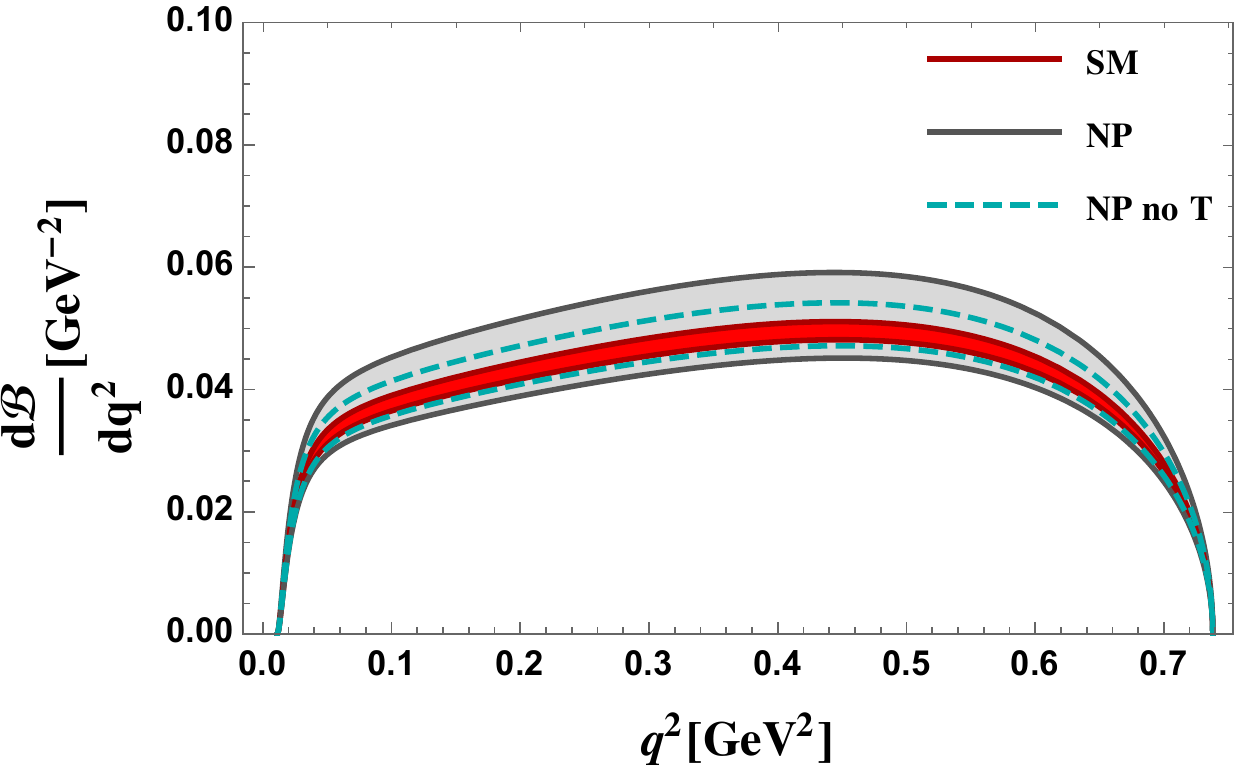} 
\caption{ \baselineskip 10pt  \small   $q^2$ spectrum of  the modes  $B_c^+ \to B_s  \mu^+ \nu_\mu$ (top) and $B_c^+ \to B_s^*  \mu^+ \nu_\mu$ (bottom). The Standard Model result (red SM band) includes the uncertainty on the form factors. The result for the full Hamiltonian Eq.~\eqref{hamil} is obtained varying the effective couplings in the quoted ranges (gray NP band). For   $B_c^+ \to B_s^*  \mu^+ \nu_\mu$ the spectrum obtained omitting the tensor operator $T$ is  also displayed (dashed  cyan lines).}\label{fig:BrBs}
\end{center}
\end{figure}

In the case of $\mu$  we describe below how  the branching fraction  changes due to the NP operators, studying also  the correlation with other observables. We notice that the $q^2$ spectrum  in  Fig.~\ref{fig:BrBs} is modified with respect to the Standard Model  when the additional operators in \eqref{hamil} are considered. The  SM prediction including the FF uncertainty  is enlarged if the  NP operators are considered, varying the  couplings  $\epsilon_i^\mu$ in their quoted ranges. However,  the shape of the spectrum is unchanged.

\begin{figure}[t]
\begin{center}
\vspace*{-1.7cm}
\hspace*{-0.5cm}
\includegraphics[width = 0.55\textwidth]{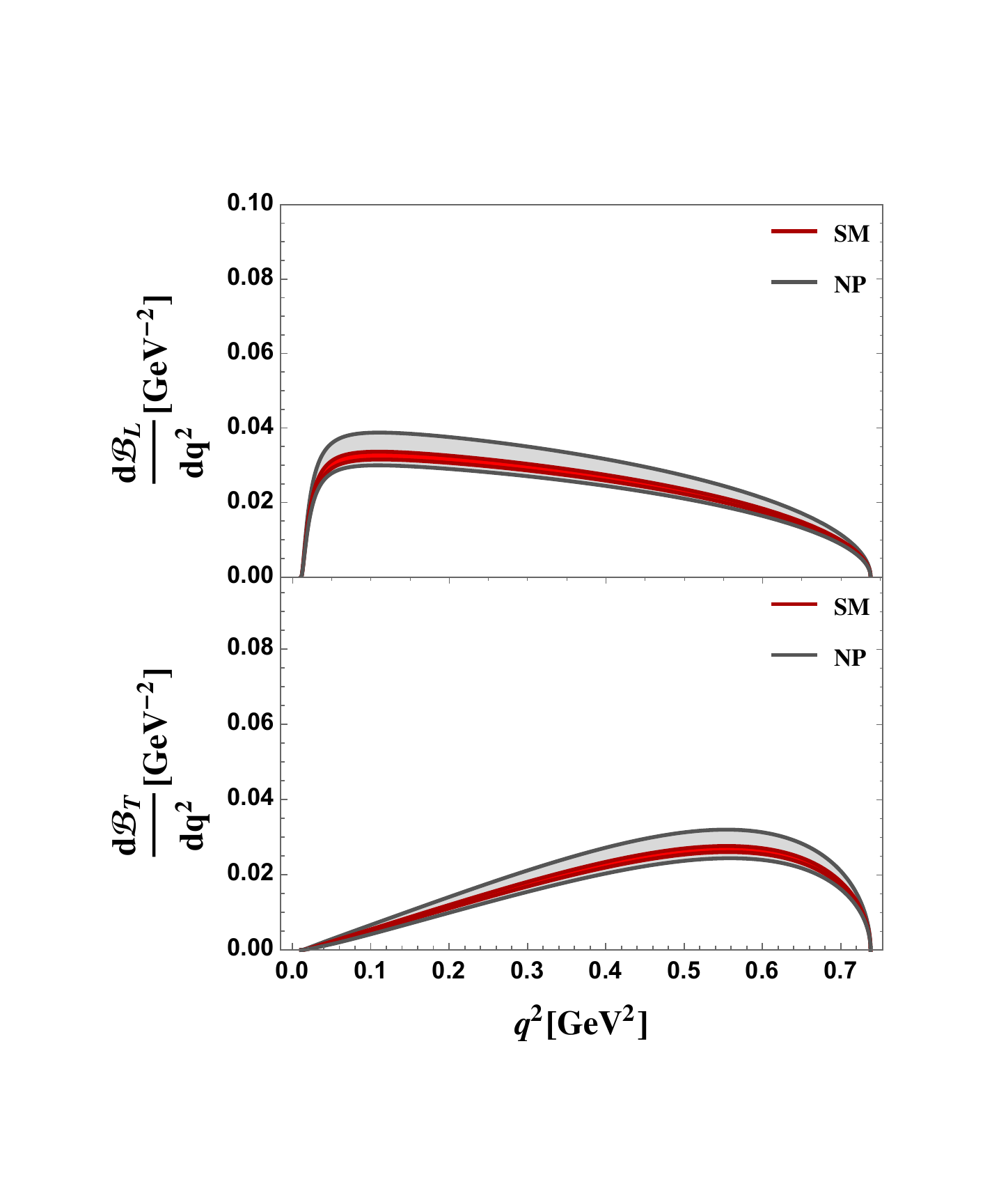} 
\vspace*{-1.7cm}
\caption{ \baselineskip 10pt  \small   $q^2$  distribution for longitudinally (top) and transversely polarized $B_s^*$ meson (bottom) in   $B_c^+ \to B_s^*  \mu^+ \nu_\mu$. The color codes are the same as in  Fig.~\ref{fig:BrBs}.  }\label{fig:BrBsstarL}
\end{center}
\end{figure}

\begin{figure}[b]
\begin{center}
\includegraphics[width = 0.48\textwidth]{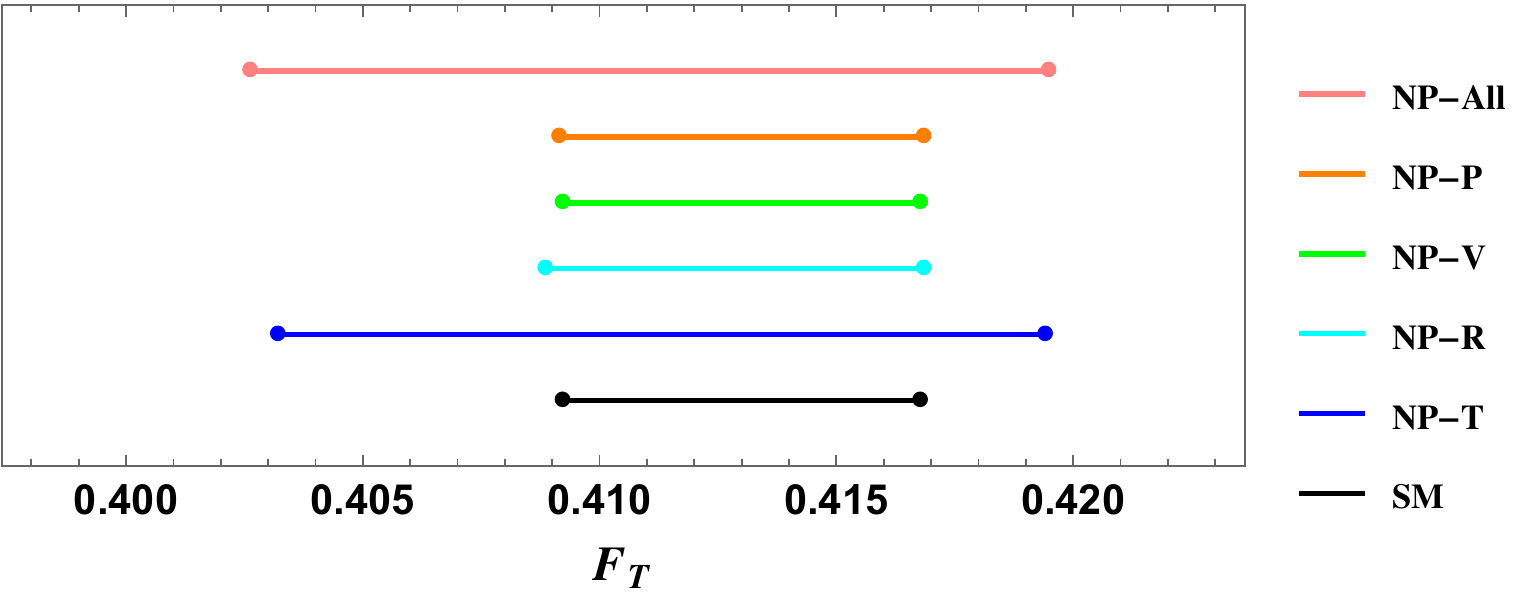} 
\caption{ \baselineskip 10pt  \small  Fraction of  transversely polarized  $B_s^*$. The lines correspond to  SM,  to  the NP operators in Eq.~\eqref{hamil}  separately considered, and to the full set of NP operators.}\label{fig:FTBsstar}
\end{center}
\end{figure}

For $B_c^+ \to B_a^* \mu^+  \nu_\mu$ ($a=s,d$),
the SM helicity amplitudes  \eqref{HampV}  can be expressed  in terms of  $\Omega_{1a}$ and $\Omega_{2a}$:
\bea
H_0&=&\sqrt{\frac{m_{B_c}}{m_{B_a^*}}}\frac{(m_{B_c}^2-m_{B_a^*}^2-q^2)}{\sqrt{q^2}}\Omega_{1a}\nn \\
&+&\frac{\lambda(m_{B_c}^2,m_{B_a^*}^2,q^2)}{2\sqrt{m_{B_c}m_{B_a^*}q^2}}\,a_0 \Omega_{2a}  \nn \\
H_\pm&=&\sqrt{\frac{m_{B_a^*}}{m_{B_c}}} \bigg(2 m_{B_c} \Omega_{1a}  \nn \\
&\mp& 
\lambda^{1/2}(m_{B_c}^2,m_{B_a^*}^2,q^2) \,a_0 \Omega_{2a} \bigg)  \\
H_t&=&-\frac{\lambda^{1/2}(m_{B_c}^2,m_{B_a^*}^2,q^2)}{2\sqrt{m_{B_c}m_{B_a^*}q^2}} 
\bigg(2m_{B_c}\Omega_{1a}\nn \\
&+&(m_{B_c}^2-m_{B_a^*}^2+q^2)\,a_0 \Omega_{2a} \bigg) , \nn
\eea
while the NP amplitudes \eqref{HampNP} read:
\bea
H_{\pm}^{NP}&= &2 \, \sqrt{\frac{m_{B^*_a}}{m_{B_c} q^2}} \nn \\
&\bigg[& \bigg(m_{B_c}^2 - m_{B^*_a}^2 + q^2 \pm \sqrt{\lambda(m_{B_c}^2,m_{B^*_a}^2,q^2)} \bigg) \, \Omega_1  \nn \\
&+& \bigg((m_{B_c} + m_{B^*_a}) \, \big((m_{B_c} - m_{B^*_a})^2 - q^2\big) \nn \\
& \pm& (m_{B_c} - m_{B^*_a}) \, \sqrt{\lambda(m_{B_c}^2,m_{B^*_a}^2,q^2)} \bigg) \, a_0 \, \Omega_2 \bigg]  \nn \\ \\
H_{L}^{NP}&=&\frac{16}{\sqrt{m_{B_c} m_{B^*_a}}} \, \bigg[ (m_{B_c}^2 + m_{B^*_a}^2 - q^2) \, \Omega_1 \nn \\
&-& m_{B^*_a} \, \big((m_{B_c} - m_{B^*_a})^2 - q^2\big) \, a_0 \, \Omega_2 \bigg] \;. \nn
\eea
For $a=s$ the SM  predictions  
\bea
{\cal B}(B_c^+ \to B_s^* \, \mu^+  \nu_\mu)&=&\,0.030 \, (1)\,\,\left(\frac{|V_{cs}|} {0.987}\right)^2 \nn \\
{\cal B}(B_c^+ \to B_s^* \, e^+  \nu_e)&=&\,0.032 \, (1)\,\,\left(\frac{|V_{cs}|} {0.987}\right)^2 
\label{BrBsstarRis}
\eea
include only the error on the form factors. For $\mu$ channel, 
the $q^2$  distribution  in Fig.~\ref{fig:BrBs} is affected by a small FF uncertainty.  In the NP extension the tensor operator   has a visible effect on  the  spectrum.
Moreover, the spectra of longitudinally and  transversely polarized $B_s^*$   in Fig.~\ref{fig:BrBsstarL}
 show that NP  mainly affects the longitudinal  $B^*_s$ polarization in the small $q^2$ region. The ratio $\dd F_T=\frac{\Gamma_T}{\Gamma_L+\Gamma_T}$, with $\Gamma_{T,L}$  the decay widths to  transversely and longitudinally polarized $B_s^*$,   is predicted in the SM: $F_T=0.413 \pm 0.004$, and remains smaller than $1/2$ when the NP operators are included, with the main effect due to the $T$ operator, as shown in Fig.~\ref{fig:FTBsstar}. 

\begin{figure}[t]
\begin{center}
\includegraphics[width = 0.41\textwidth]{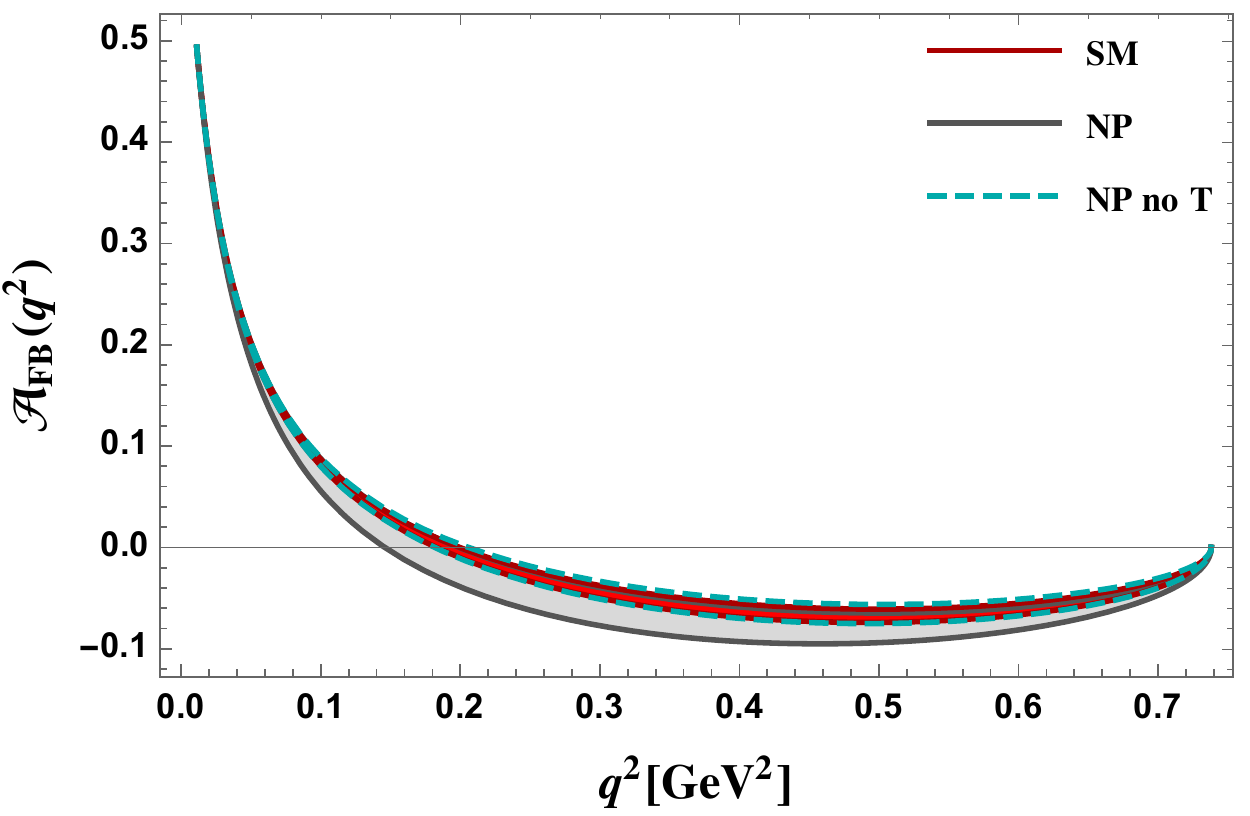}  
\caption{ \baselineskip 10pt  \small $q^2$-dependent  forward-backward lepton asymmetry in  $B_c^+ \to B_s^*  \mu^+ \nu_\mu$. The red band corresponds to SM, the gray band to the full Hamiltonian \eqref{hamil}. The region obtained excluding the tensor operator $T$ is  indicated by the  dashed cyan lines.}\label{fig:AFB}
\end{center}
\end{figure}

The $q^2$-dependent forward-backward (FB) lepton asymmetry
\bea
&&
{\cal A_{FB}}(q^2)=\left({\displaystyle{\frac{d \Gamma}{dq^2}}} \right)^{-1}\times \label{AFB} \\
&&
 \left[\int_0^1 \, d\cos \, \theta \, \displaystyle{\frac{d^2 \Gamma}{dq^2 d\cos \, \theta}} -\int_{-1}^0 \, d\cos \, \theta \, \displaystyle{\frac{d^2 \Gamma}{dq^2 d\cos \, \theta}} \right] \,\,\, \nn
\eea
is affected by a small uncertainty in the SM  (Fig.~\ref{fig:AFB}). The asymmetry has a zero precisely determined at $q_0^2 \simeq 0.1905\,(5)$ GeV$^2$. This observable is  particular sensitive to the  tensor operator:   indeed, as shown in Fig.~\ref{fig:AFB},  excluding  this operator the asymmetry in NP practically coincides with SM. When all the operators in the extended Hamiltonian are considered the position of the zero is in the range $q_0^2 \in [0.149,\,0.208]$ GeV$^2$. 

\begin{figure}[b]
\begin{center}
\includegraphics[width = 0.42\textwidth]{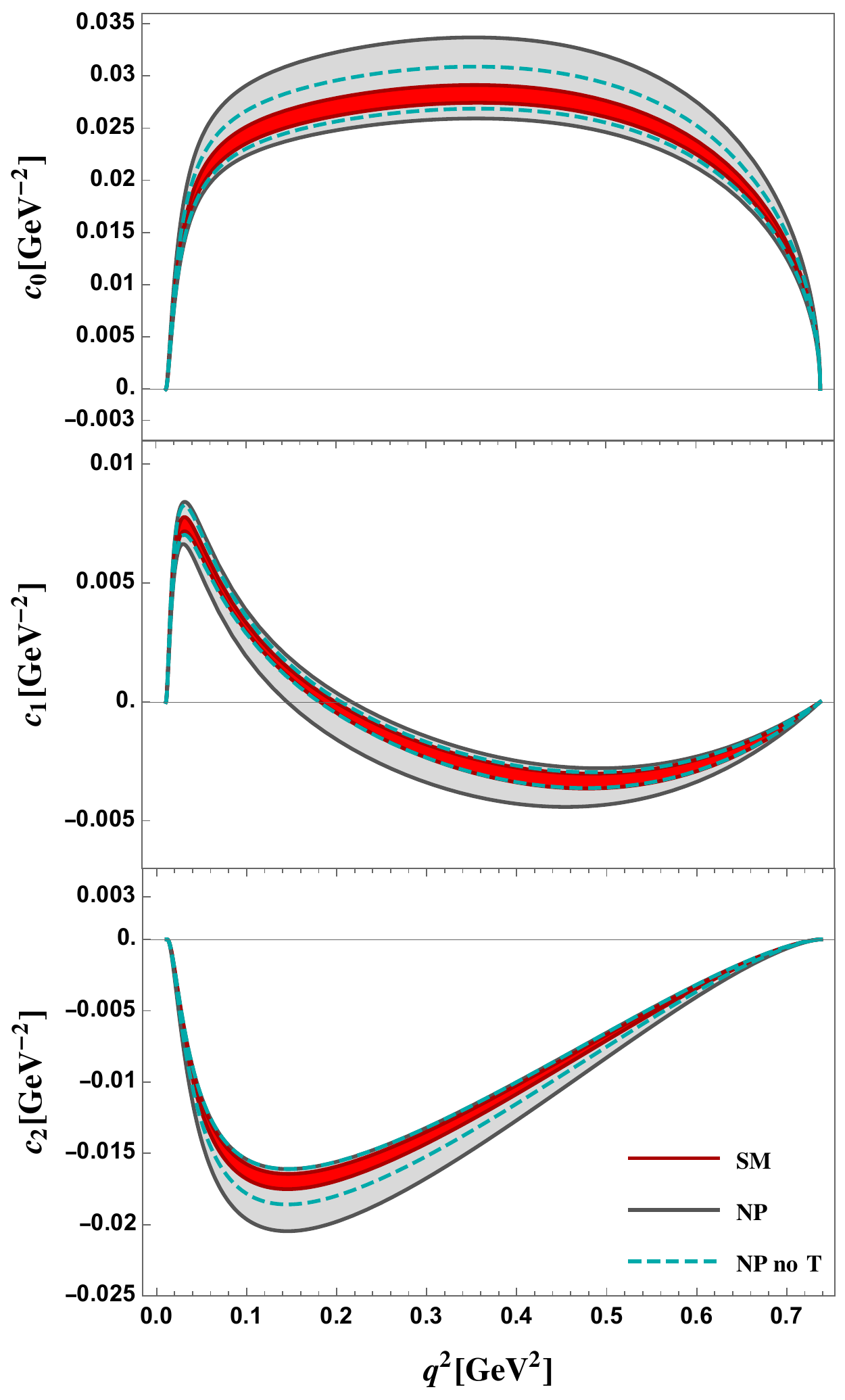}  
\caption{ \baselineskip 10pt  \small Coefficients $c_{0,1,2}$ in  Eq.~\eqref{eq:c012} for $B_c \to B^*_s \mu^+ \nu_\mu$. The color codes are  the same as in  Fig.~\ref{fig:AFB}. }\label{fig:a0a1a2Bsstar}
\end{center}
\end{figure}

The effects of the new operators  can also be observed in the coefficients $c_{0,1,2}$ defined in the expression \cite{Penalva:2020xup,Penalva:2020ftd}
\be
\frac{d {\cal B}(B_c^+ \to B_s^* \, \mu^+  \nu_\mu)}{dq^2 d \cos \theta}= c_0 +c_1 \cos \theta + c_2 \cos^2 \theta \,\, , \label{eq:c012}
\ee
as  shown in Fig.~\ref{fig:a0a1a2Bsstar}.

Interesting information is encoded in  the correlations between  the various observables in the  decay modes to the pseudoscalar and vector meson.  We analyze them in turn, neglecting the common FF uncertainties, considering the SM,   each  NP operator and all operators together. Since the scalar and pseudoscalar operators have a minor impact on the results,   we do not discuss them individually. 
\begin{figure}[b]
\begin{center}
\includegraphics[width = 0.45\textwidth]{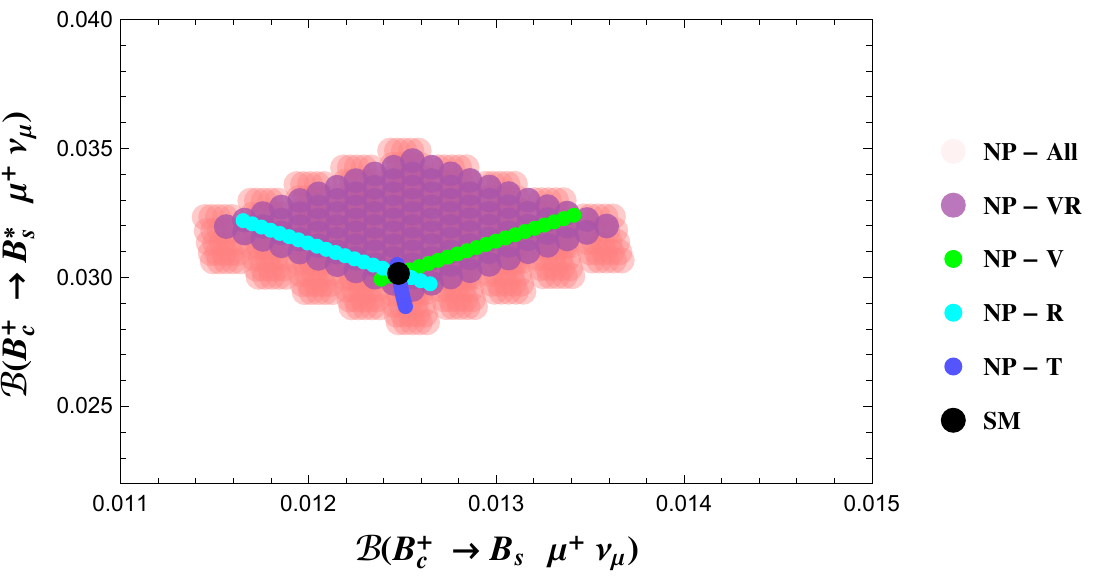}  
\caption{ \baselineskip 10pt  \small Correlation between the branching fractions   ${\cal B}(B_c^+ \to B_s  \mu^+ \nu_\mu$) and  ${\cal B} (B_c^+ \to B_s^*  \mu^+ \nu_\mu$) in SM (black dot) and for the NP operators in Eq.~\eqref{hamil}. The regions labeled $VR$, $V$, $R$ and $T$  are obtained varying  separately the  coefficients of the corresponding operators in their quoted ranges. The NP-All region refers to  the full set of operators in \eqref{hamil}.   }\label{fig:Brscorr}
\end{center}
\end{figure}

\begin{figure}[t]
\begin{center}
\includegraphics[width = 0.45\textwidth]{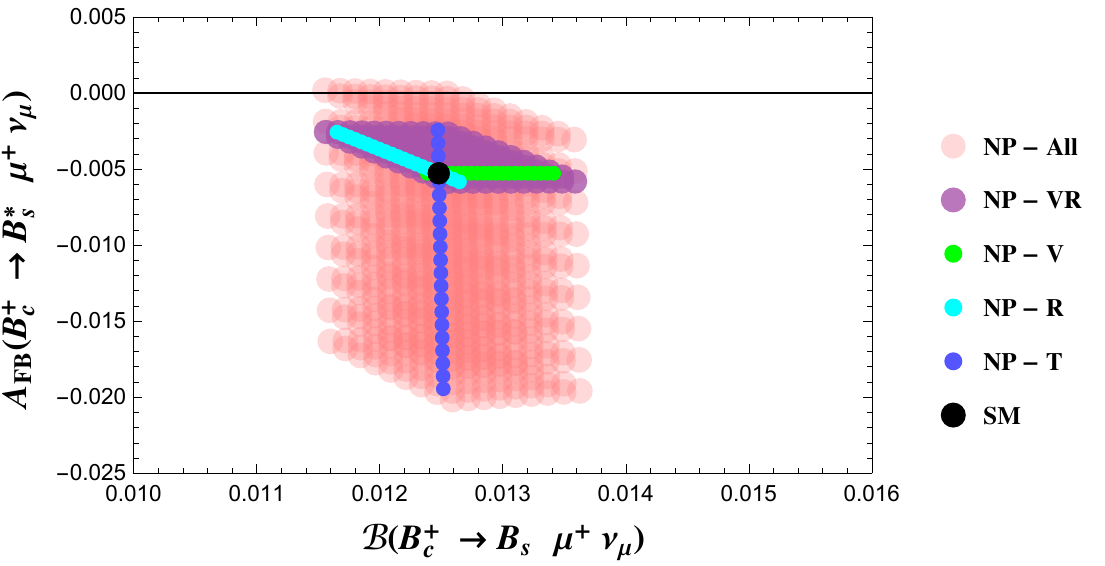} \\
\includegraphics[width = 0.45\textwidth]{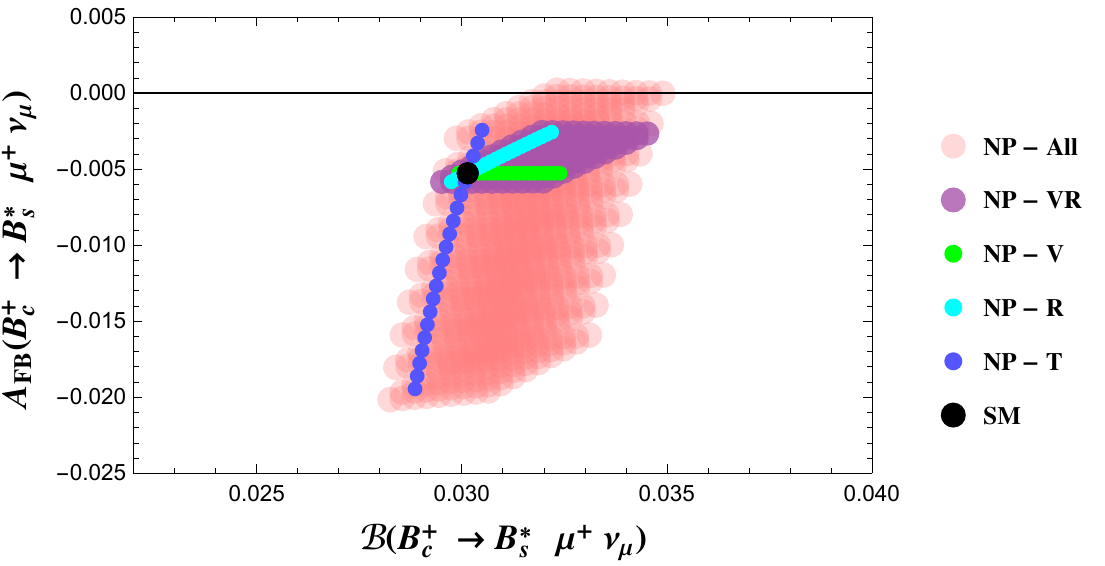} 
\caption{ \baselineskip 10pt  \small Correlations between the integrated forward-backward lepton asymmetry $A_{FB}$ in $B_c^+ \to B_s^*  \mu^+  \nu_\mu$,  defined in Eq.~\eqref{eq:AFB}, with  ${\cal B}(B_c^+ \to B_s \, \mu^+  \nu_\mu)$ (top) and ${\cal B}(B_c^+ \to B_s^*  \mu^+  \nu_\mu)$ (bottom panel).  The color codes are the same as in  Fig.~\ref{fig:Brscorr}.}\label{fig:BrAFBcorr}
\end{center}
\end{figure}

Fig.~\ref{fig:Brscorr} shows  the correlation between the branching fractions of the pseudoscalar and vector modes  ${\cal B}(B_c^+ \to B_s \, \mu^+  \nu_\mu)$ and ${\cal B}(B_c^+ \to B_s^*\, \mu^+  \nu_\mu)$. The SM point  corresponds to the central values  in Eqs.~(\ref{BrBsRis}) and (\ref{BrBsstarRis}).  When all NP operators are considered the enlarged  (pink) region is obtained. Anticorrelation between the branching fractions  is found when the $R$ operator is considered.  Increasing  $\epsilon_V^\mu$  produces a positive correlation between the two observables.   The tensor operator $T$ can allow a reduction of   ${\cal B}(B_c^+ \to B_s^*  \,\mu^+  \nu_\mu)$  with respect to  SM.
Structured patterns are found in  the correlations of the branching fractions  ${\cal B}(B_c^+ \to B_s \, \mu^+  \nu_\mu)$ and ${\cal B}(B_c^+ \to B_s^* \, \mu^+  \nu_\mu)$  with  the integrated FB lepton asymmetry in the $B^*_s$ mode
\be
A_{FB}=\int_{q^2_{min}}^{q^2_{max}} dq^2 \, {\cal A_{FB}}(q^2)\,,   \label{eq:AFB}
\ee
as shown  in Fig.~\ref{fig:BrAFBcorr}. Varying the $R$ and $V$ coefficients produces   anticorrelations in case of the $B_s$ channel,  same sign correlation in case of $B_s^*$. 
 The tensor operator  results in  a mild anticorrelation in the $B_s^*$ case. The combined analysis of all  observables can allow to isolate  the signature of the different  NP operators.

\subsection{$B_c^+ \to B_d \ell^+  \nu_\ell$ and $B_c^+ \to B_d^*(\to B_d \gamma) \ell^+  \nu_\ell$ }
The  $c \to d$ semileptonic $B_c$ modes also give access to relevant information.  The SM expectations
\bea
{\cal B}(B_c^+ \to B_d \,\mu^+  \nu_\mu)&=&8.3 \, (5) \times 10^{-4}\, \left(\frac{|V_{cd}|}{0.221}\right)^2  \nn \\
{\cal B}(B_c^+ \to B_d \,e^+  \nu_e)&=&8.7 \, (5) \times 10^{-4}\, \left(\frac{|V_{cd}|}{0.221}\right)^2 \,\,\,\,\,\,\,\,\,
\,\label{BrBdRis}
\eea
derive from the form factors in \cite{Cooper:2020wnj}. The quoted errors are only due to the FF uncertainty.
The corresponding predictions  for $B_c^+ \to B_d^*  \,\bar \ell \nu_\ell$ in SM are
\bea
{\cal B}(B_c^+ \to B_d^* \,\mu^+  \nu_\mu)&=&20 \,(1) \times 10^{-4}\, \left(\frac{|V_{cd}|}{0.221}\right)^2   \nn \\
{\cal B}(B_c^+ \to B_d^* \, e^+  \nu_e)&=&21 \,(1) \times 10^{-4}\, \left(\frac{|V_{cd}|}{0.221}\right)^2. \,\,\,\,\,\,\,\,\, \label{BrBdstarRis}
\eea

\begin{figure}[t]
\begin{center}
\includegraphics[width = 0.45\textwidth]{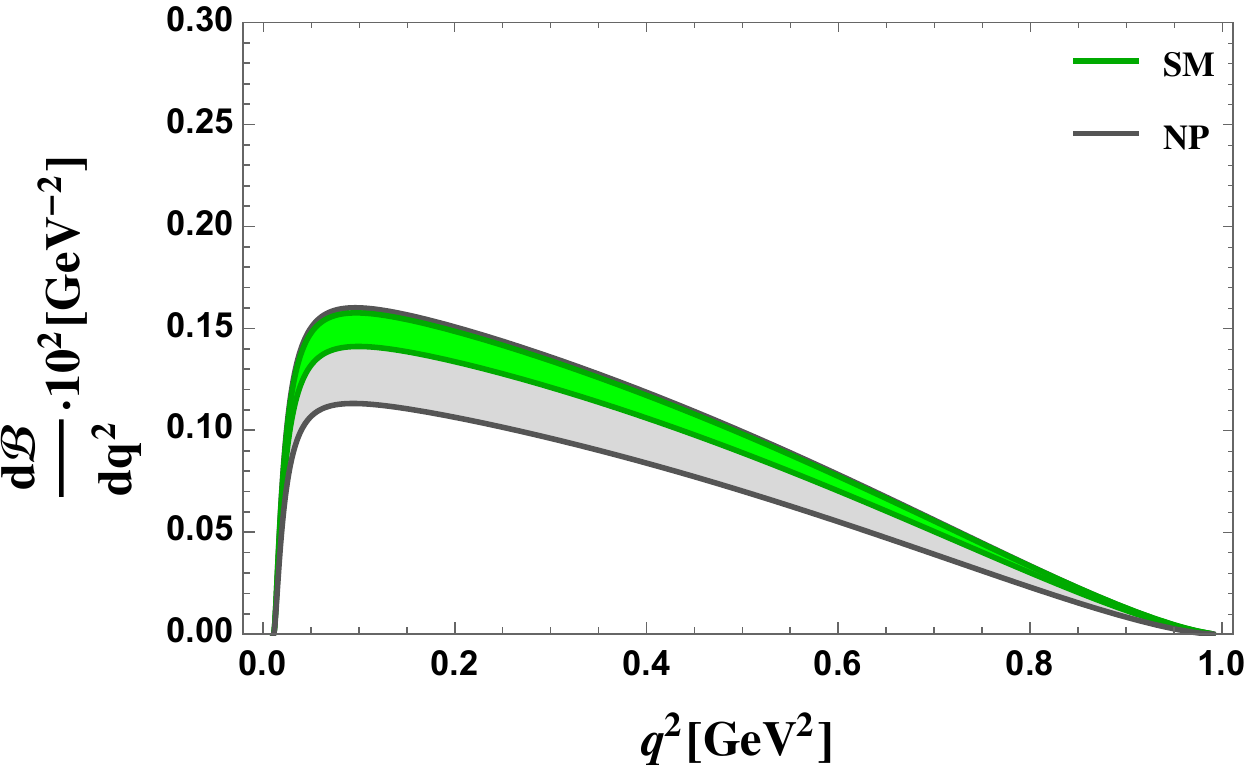} \\
\includegraphics[width = 0.45\textwidth]{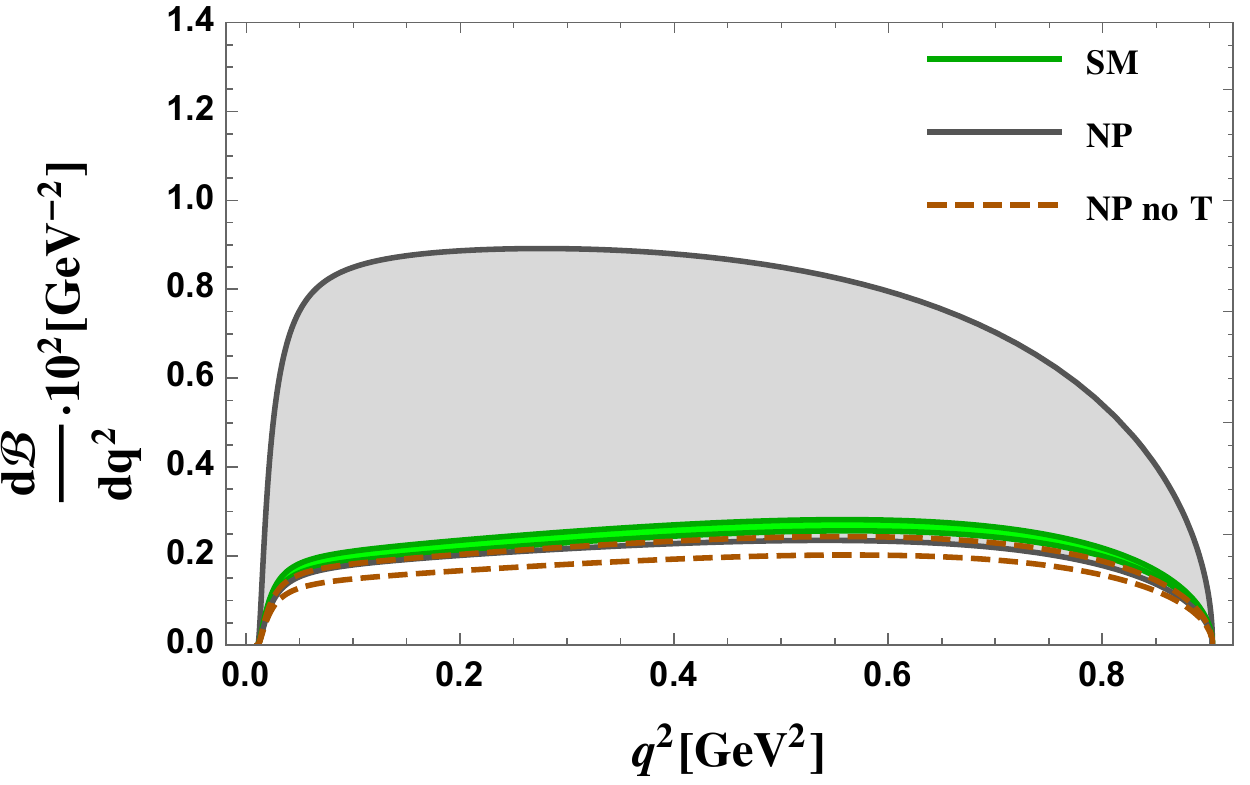} 
\caption{ \baselineskip 10pt  \small   $q^2$ spectrum of  the  modes $B_c^+ \to B_d  \, \mu^+ \nu_\mu$ (top) and $B_c^+ \to B_d^*  \, \mu^+ \nu_\mu$ (bottom). The Standard Model results (green SM band) include the uncertainty on the form factors. The spectra for the full Hamiltonian in Eq.~\eqref{hamil} are obtained varying the effective couplings in their quoted ranges (gray NP band). For   $B_c^+ \to B_d^*  \, \mu^+ \nu_\mu$ the spectrum obtained omitting the tensor operator $T$ is also shown (dashed orange lines). }\label{fig:BrBd}
\end{center}
\end{figure}

\begin{figure}[t]
\begin{center}
\includegraphics[width = 0.41\textwidth]{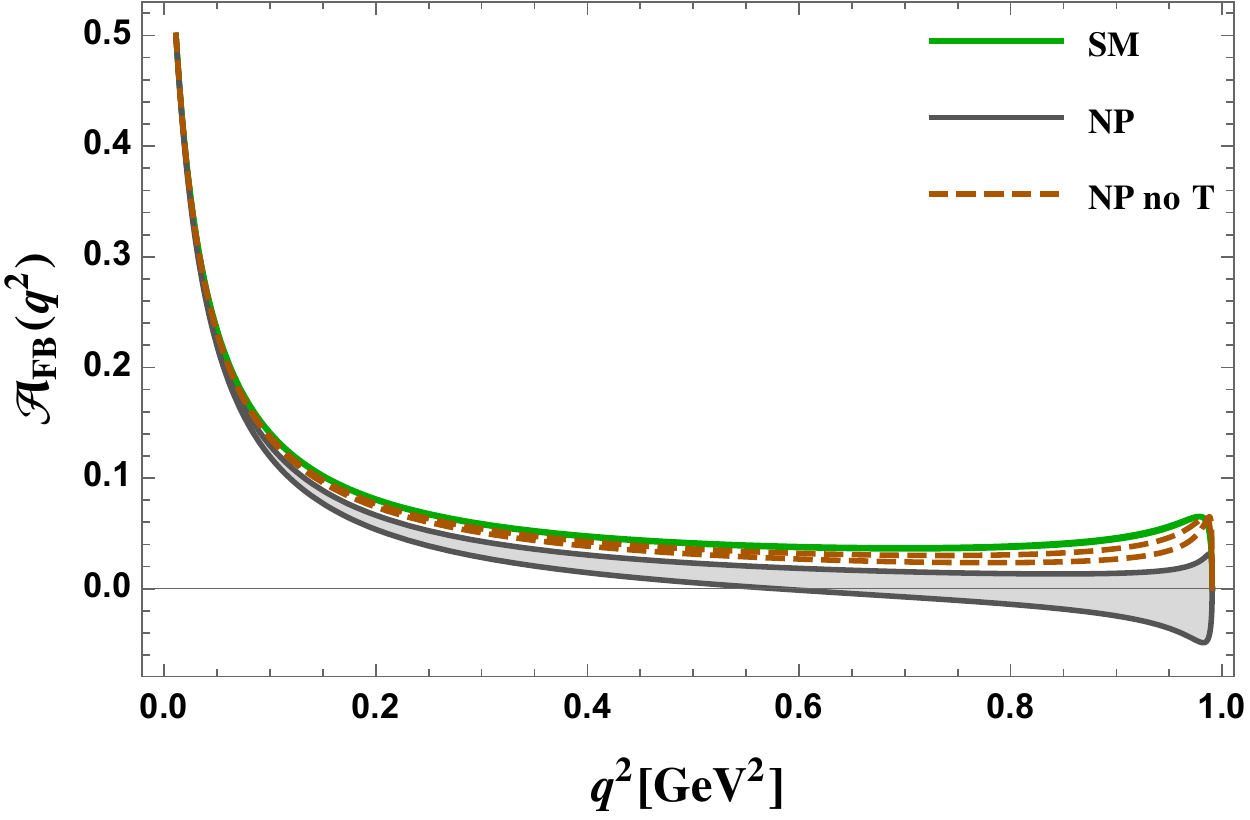} 
\caption{ \baselineskip 10pt  \small  $q^2$-dependent forward-backward  lepton asymmetry  in  $B_c^+ \to B_d  \, \mu^+ \nu_\mu$. The green line corresponds to SM, the gray band is obtained for the Hamiltonian \eqref{hamil}. The dashed orange lines are  obtained excluding the tensor operator $T$. }\label{fig:AFBctoBd}
\end{center}
\end{figure}

For the $\mu$ channel, the impact of the NP operators in the   decay distributions is shown in  Fig.~\ref{fig:BrBd}. The spectra  in SM   are affected by a  small FF uncertainty.  Including the NP operators sizably enlarges the spectrum of the pseudoscalar mode. The forward-backward asymmetry
Eq.~\eqref{AFB} for the pseudoscalar mode shows deviations from the SM expectation mainly due to  the tensor operator, Fig.~\ref{fig:AFBctoBd}.

Large effects are allowed  in  $B_c^+ \to B_d^*(\to B_d \gamma) \ell^+  \nu_\ell$: this is due to the contribution of the tensor operator, that overwhelms the other ones if the coefficient $\epsilon_T^\mu$ is varied in  the parameter space bound in \cite{Becirevic:2020rzi} using $D$ meson decays.
 
\begin{figure}[b]
\begin{center}
\vspace*{-1.9cm}
\hspace*{-0.5cm}
\includegraphics[width = 0.51\textwidth]{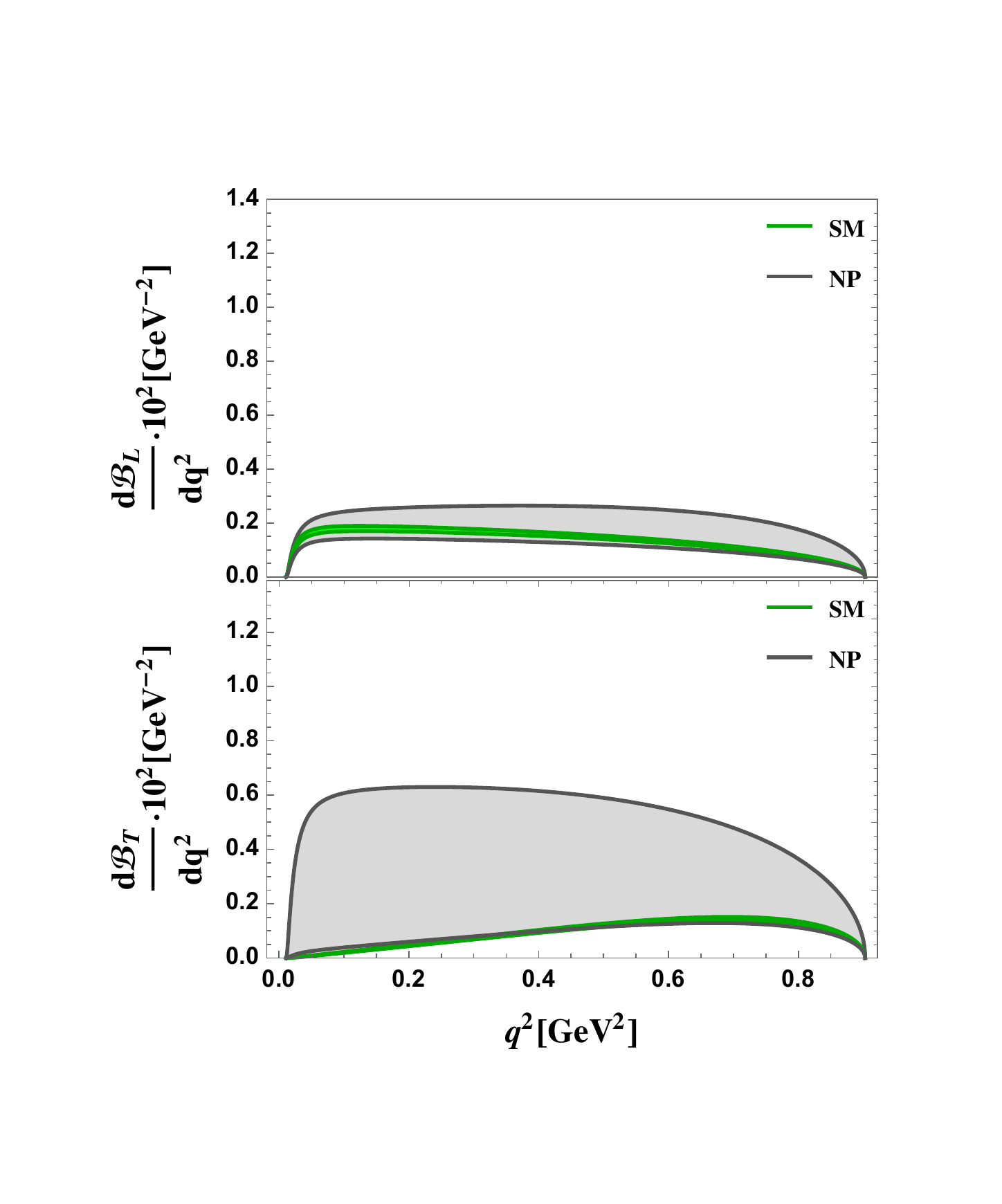} 
\vspace*{-1.cm}
\caption{ \baselineskip 10pt  \small   $q^2$  distribution of longitudinally (top) and transversely polarized $B_d^*$ (bottom) in   $B_c^+ \to B_d^* \, \mu^+ \nu_\mu$.  The color codes are  as in Fig.~\ref{fig:BrBd}. }\label{fig:BrBdstarL}
\end{center}
\end{figure}

\begin{figure}[t]
\begin{center}
\includegraphics[width = 0.48\textwidth]{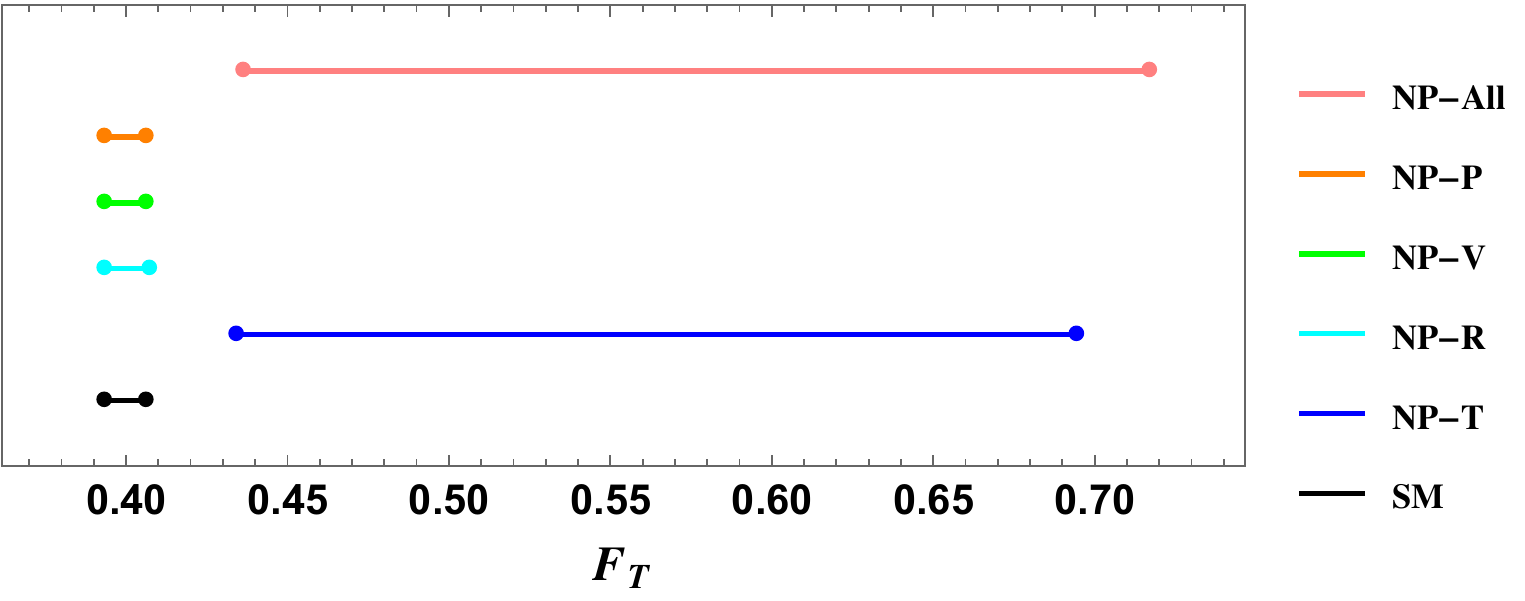} 
\caption{ \baselineskip 10pt  \small   Fraction of  transversely polarized  $B_d^*$. The lines correspond to   the SM,  to  the NP operators in Eq.~\eqref{hamil}  separately considered, and to the full set of NP operators.  }\label{figFTBdstar}
\end{center}
\end{figure}

The distributions of longitudinally and transversely polarized $B_d^*$, Fig.~\ref{fig:BrBdstarL}, show  that the tensor operator can sizably affect the transverse distribution. In SM   the integrated width to longitudinal $B^*_d$  is larger than to the transverse one, as shown in  Fig.~\ref{figFTBdstar}. The  tensor operator  can reverse such a hierachy.

\begin{figure}[b]
\begin{center}
\includegraphics[width = 0.41\textwidth]{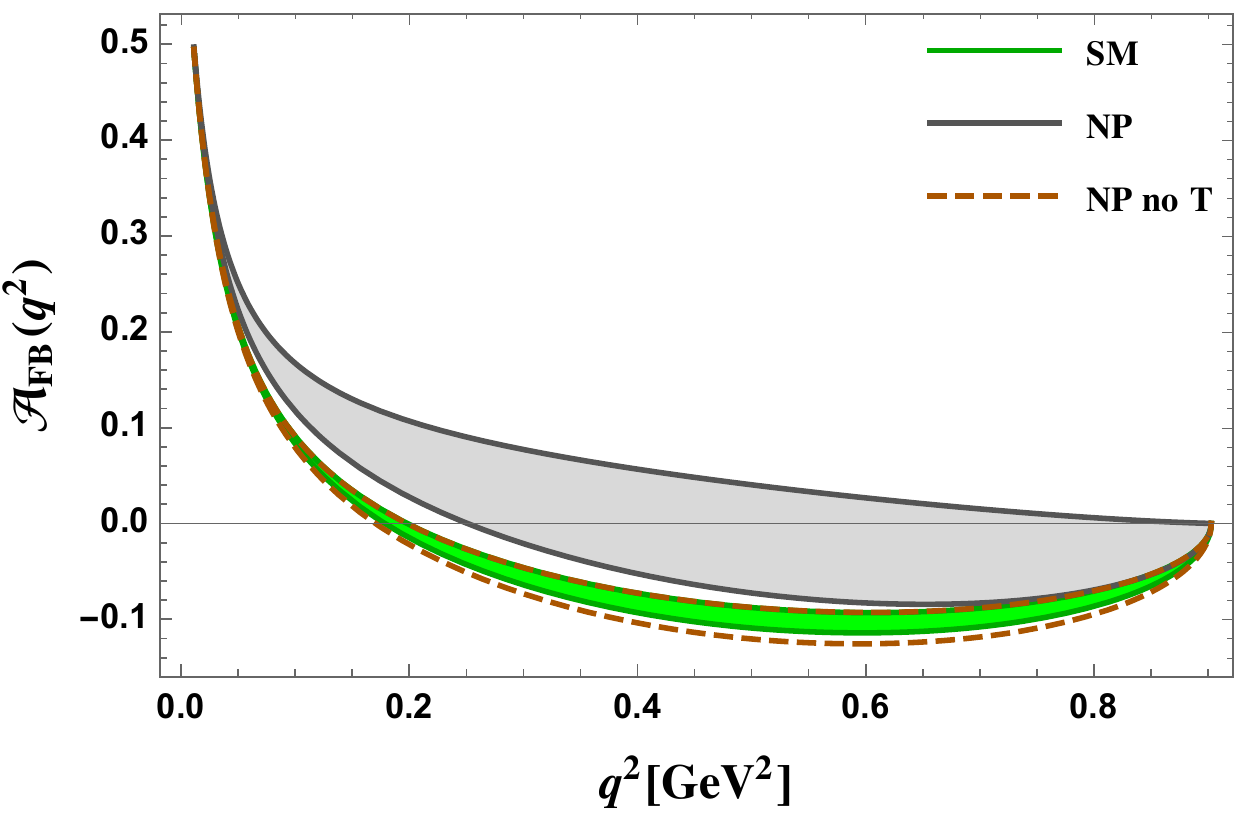} 
\caption{ \baselineskip 10pt  \small  $q^2$-dependent forward-backward  lepton asymmetry  in  $B_c^+ \to B_d^*  \, \mu^+ \nu_\mu$. The green band corresponds to SM, the gray one is obtained for the Hamiltonian \eqref{hamil}. The region obtained excluding the tensor operator $T$ (dashed orange lines) is also displayed. }\label{fig:AFBd}
\end{center}
\end{figure}

\begin{figure}[t]
\begin{center}
\includegraphics[width = 0.42\textwidth]{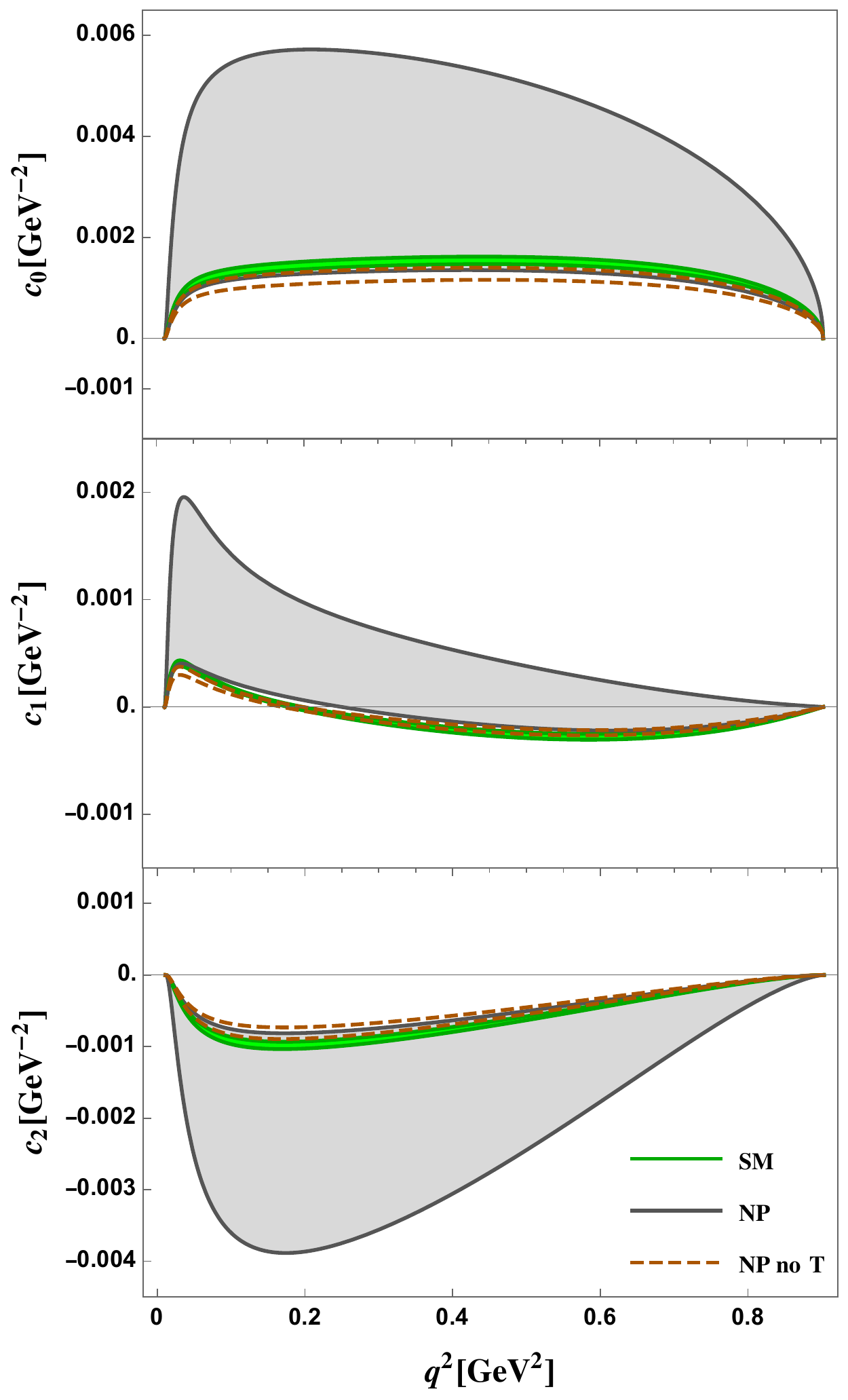}  
\caption{ \baselineskip 10pt  \small Coefficients $c_{0,1,2}$ in Eq.~\eqref{eq:c012} for $B_c \to B^*_d \mu^+ \nu_\mu$. The color codes are   as in  Fig.~\ref{fig:AFBd}. }\label{fig:a0a1a2Bdstar}
\end{center}
\end{figure}

Also the $q^2$-dependent forward-backward  lepton asymmetry  shows this effect, as  seen in Fig.~\ref{fig:AFBd}. The inclusion of the tensor  operator produces a zero for the ${\cal A_{FB}}$  distribution in the  range $q_0^2 \in [0.27\, {\rm GeV}^2 ,\,q^2_{max}]$, while  in the SM  $q_0^2= 0.188(1)$ GeV$^2$ is expected. The position of the zero of ${\cal A_{FB}}(q^2)$  has a remarkable discriminating power of  NP operators. 
 The effects of the new operators on the coefficients defined in \eqref{eq:c012} are shown in Fig.~\ref{fig:a0a1a2Bdstar}.


\begin{figure}[t]
\begin{center}
\includegraphics[width = 0.45\textwidth]{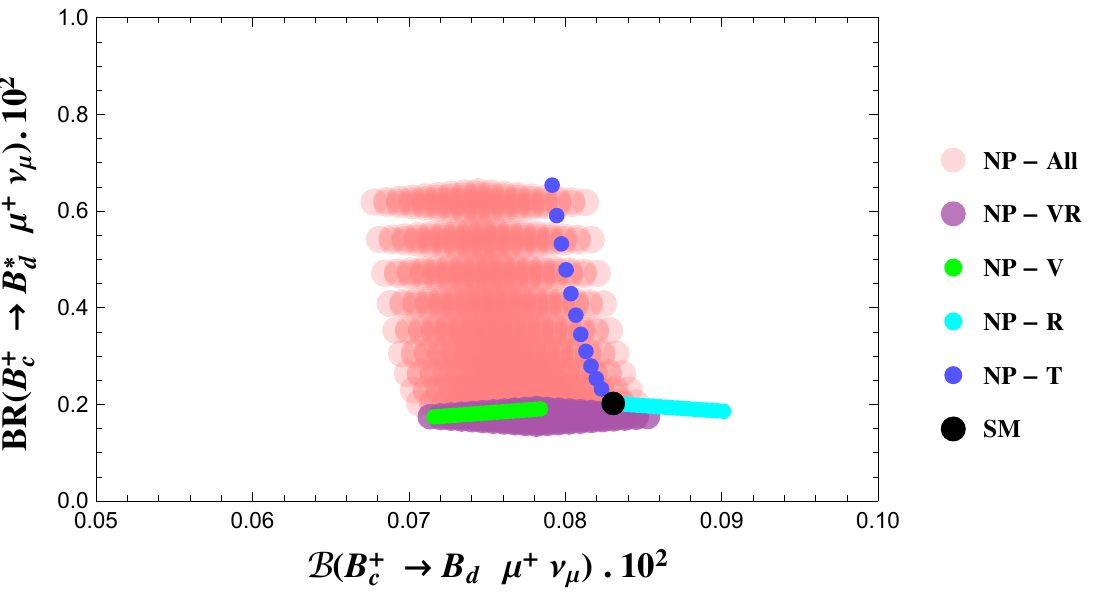} 
\caption{ \baselineskip 10pt  \small  Correlation between the branching fractions   ${\cal B}(B_c^+ \to B_d  \mu^+ \nu_\mu$) and  ${\cal B} (B_c^+ \to B_d^*  \mu^+ \nu_\mu$) in SM (black dot) and considering  the NP operators in Eq.~\eqref{hamil}. The regions labeled $VR$, $V$, $R$ and $T$  are obtained varying  separately the  coefficients of the corresponding operators in their quoted ranges. The NP-All region refers to the full set of operators in \eqref{hamil}.    }\label{fig:Brdcorr}
\end{center}
\end{figure}

\begin{figure}[b]
\begin{center}
\includegraphics[width = 0.45\textwidth]{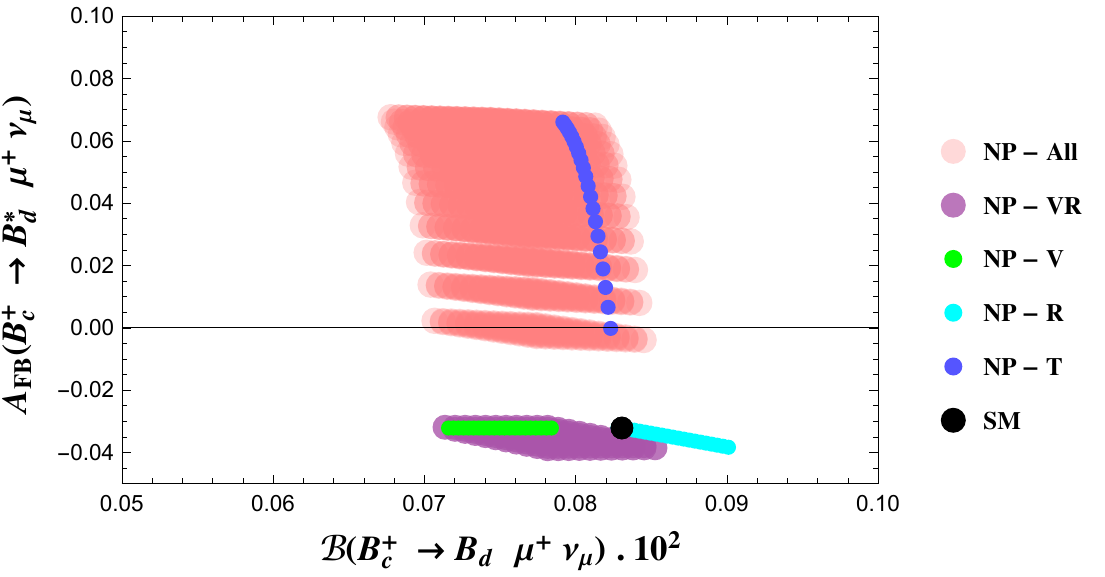} \\
\includegraphics[width = 0.45\textwidth]{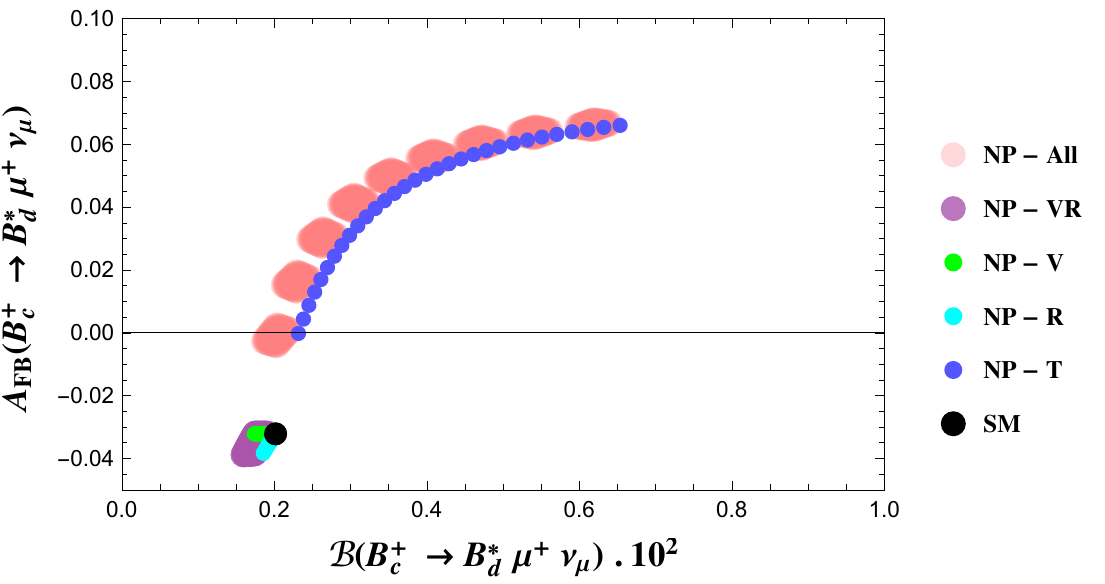} 
\caption{ \baselineskip 10pt  \small Correlations between the integrated forward-backward lepton asymmetry  $A_{FB}$ in $B_c^+ \to B_d^*  \mu^+  \nu_\mu$,  defined in Eq.~\eqref{eq:AFB},  with  ${\cal B}(B_c^+ \to B_d \, \mu^+  \nu_\mu)$ (top) and ${\cal B}(B_c^+ \to B_d^*  \mu^+  \nu_\mu)$ (bottom panel).  The color codes are the same as in Fig.~\ref{fig:Brdcorr}.  }\label{fig:BrdAFBcorr}
\end{center}
\end{figure}

The correlation plots   in Figs.~\ref{fig:Brdcorr} and \ref{fig:BrdAFBcorr} give access to  other information. The branching factions ${\cal B}(B_c^+ \to B_d^* \,  \mu^+ \nu_\mu) $ and ${\cal B}(B_c^+ \to B_d \,  \mu^+ \nu_\mu) $ are sizably affected by the NP contributions. 
The $R$  operator  anti-correlates the decay widths of the pseudoscalar and vector modes,  while the $V$ contribution results in a positive correlation. In particular,  ${\cal B}(B_c^+ \to B_d^* \,  \mu^+ \nu_\mu) $ increases with respect to  SM  if   $R$ is included, and decreases considering only  $V$. However, the main effect is due to the  tensor  operator that strongly enhances ${\cal B}(B_c^+ \to B_d \,  \mu^+ \nu_\mu)$   if its coefficient is varied in the range quoted  in  \cite{Becirevic:2020rzi}. Such a macroscopic effect  on the one hand requires to further   scrutinize  the bounds from the $D$ meson decays,  on the other hand shows the relevance of the  $B_c$ modes in the search of BSM signals. This is confirmed by the  correlations between  the integrated forward-backward lepton  asymmetry $A_{FB}$ and the branching fractions of the pseudoscalar and vector modes. As shown in   Fig.~\ref{fig:BrdAFBcorr}, the integrated $A_{FB}$, that in SM is predicted to be negative,  is anti-correlated with ${\cal B}(B_c^+ \to B_d  \,\mu^+ \nu_\mu)$ mainly due to the tensor operator. $A_{FB}$ can become positive in the allowed range for the coefficient of such an operator, an interesting experimental signature. On the other hand, $A_{FB}$ and ${\cal B}(B_c^+ \to B_d^*   \,\mu^+ \nu_\mu)$ are positively correlated, and the enhancement of the branching fraction closely follows the enhancement of $A_{FB}$ obtained varying the  coefficient of the tensor operator.

\section{Conclusions}
The semileptonic $B_c$ decays induced by the $c \to s,d$ transitions play an interesting role  in SM and  in the search of BSM  effects analogous to the ones emerging in $B$ decays. The heavy quark spin symmetry has allowed to analyze the full phenomenology of such decays using two nonperturbative  form factors  obtained by lattice QCD. The assessment of the role of the symmetry-breaking terms requires additional nonperturbative information, namely some other form factor in few points of the kinematical range. We have studied several significant observables in these decay modes, together with  the effects and their correlations of the SM extension involving  dimension-6 operators and left-handed neutrinos.  

On the basis of
the available information on semileptonic  $D$ decays we have found that sizable deviations from SM are allowed in $B_c^+ \to B_d^* \,  \mu^+ \nu_\mu$. Of particular interest are the correlations 
of the effects of the  NP operators in the various observables, that can be used to pin-down the single contributions. For example, the branching fractions of the pseudoscalar and vector modes are positively or negatively correlated if the $R$ or $V$ contributions are considered. Other correlations involve the integrated FB lepton asymmetry, in particular the effect of the tensor operator in the $B_c^+ \to B_d^* \,  \mu^+ \nu_\mu$ mode correlated to the branching  fraction. The position of the zero in the FB lepton distribution, as well as the fraction of longitudinally vs transversely polarized final vector mesons constitute other observables  worth to measure.

\section{Acknowledgements} 
 We thank D. Be\v{c}irevi\'c, F. Jaffredo,  A. Pe\~nuelas  and O. Sumensari for communications about Ref.~\cite{Becirevic:2020rzi}. 
This study has been  carried out within the INFN project (Iniziativa Specifica) QFT-HEP.

\appendix
\numberwithin{equation}{section}
\section{Hadronic matrix elements and  form factors in SM and NP}\label{app:A}
We use the standard parametrization of the hadronic $B_c \to P, V$ matrix elements in terms of form factors, with $P$ a pseudoscalar and $V$ a vector meson.
The $B_c \to P$  matrix elements of the vector  $\bar q \gamma_\mu Q$  current, of the scalar density 
$\bar q Q$, and of the tensor $\bar q \sigma_{\mu \nu} Q$  and $\bar q \sigma_{\mu \nu} \gamma_5 Q$ currents are parametrized  as:
\bea
&&\langle P(p^\prime)| {\bar q} \gamma_\mu Q| {B_c}(p) \rangle = \nn \\
&&\qq  f_+^{B_c \to P}(q^2) \Big(p_\mu+p_\mu^\prime  - \frac{m_{B_c}^2-m_P^2}{q^2} q_\mu\Big) \nn \\
&&\qq + \,f_0^{B_c \to P}(q^2)\frac{m_{B_c}^2-m_P^2}{q^2} q_\mu \,\,,  \nn  \\ 
&&\langle P(p^\prime)| {\bar q} Q| {B_c}(p) \rangle = f_S^{B_c \to P}(q^2) \,\,, \label{BctoP} \\
&&\langle P(p^\prime)| {\bar q} \sigma_{\mu \nu }Q| B_c(p) \rangle = -i \frac{2 f_T^{B_c \to P}(q^2)}{m_{B_c}+m_P} \big(p_\mu p_\nu^\prime-p_\nu p^\prime_\mu \big) \,\,, \nn \\ 
&&\langle P(p^\prime)| {\bar q} \sigma_{\mu \nu }\gamma_5 Q| { B_c}(p) \rangle =- \frac{2 f_T^{B_c \to P}(q^2)}{m_{B_c}+m_P} \epsilon_{\mu \nu \alpha \beta} \, p^\alpha p^{\prime \beta}, \nn 
\eea
with $\epsilon^{0123}=+1$. The condition $f_+^{B_c \to P}(0)=f_0^{B_c \to P}(0)$ holds.
Moreover, one has  $f_S^{B_c \to P}(q^2)=\displaystyle \frac{m_{B_c}^2-m_P^2}{m_Q-m_q}f_0^{B_c \to P}(q^2)$ in terms of the   quark  masses $m_Q$  and $m_q$.

\noindent The $B_c \to V$ matrix elements are parametrized as:
\bea
&&\langle V(p^\prime,\epsilon)|{\bar q} \gamma_\mu Q| {B_c}(p) \rangle = 
- {2 V^{B_c \to V}(q^2) \over m_{B_c}+m_V} i \epsilon_{\mu \nu \alpha \beta} \epsilon^{*\nu}  p^\alpha p^{\prime \beta}, \nn 
\eea
\bea
&&\langle V(p^\prime,\epsilon)|{\bar q} \gamma_\mu\gamma_5 Q| {B_c}(p) \rangle = \nn \\
&&  (m_{B_c}+m_V) \Big( \epsilon^*_\mu -{(\epsilon^* \cdot q) \over q^2} q_\mu \Big) A_1^{B_c \to V}(q^2) \nn\\
&& - {(\epsilon^* \cdot q) \over  m_{B_c}+m_V} \Big( (p+p^\prime)_\mu -{m_{B_c}^2-m_V^2 \over q^2} q_\mu \Big) A_2^{B_c \to V}(q^2) \nn \\
&& + (\epsilon^* \cdot q){2 m_V \over q^2} q_\mu A_0^{B_c \to V}(q^2),  \nn \\
&&\langle V(p^\prime,\epsilon)|{\bar q} \gamma_5 Q| {B_c}(p) \rangle =-\frac{2 m_V}{m_Q+m_q} (\epsilon^* \cdot q) A_0^{B_c \to V}(q^2), \nn \\
&&\langle V(p^\prime,\epsilon)|{\bar q} \sigma_{\mu \nu} Q| { B_c}(p) \rangle =\nn \\
&&\qq T_0^{B_c \to V}(q^2) {\epsilon^* \cdot q \over (m_{B_c}+ m_V)^2} \epsilon_{\mu \nu \alpha \beta} p^\alpha p^{\prime \beta}\nn \\
&&\qq +T_1^{B_c \to V}(q^2) \epsilon_{\mu \nu \alpha \beta} p^\alpha \epsilon^{*\beta} \nn \\
&&\qq + T_2^{B_c \to V}(q^2) \epsilon_{\mu \nu \alpha \beta} p^{\prime \alpha} \epsilon^{*\beta}, 
\label{BctoV} \\ 
&&\langle V(p^\prime,\epsilon)|{\bar q} \sigma_{\mu \nu}\gamma_5 Q| { B_c}(p) \rangle = \nn \\
&&\qq i\, T_0^{B_c \to V}(q^2) {\epsilon^* \cdot q \over (m_{B_c}+ m_V)^2} (p_\mu p^\prime_\nu-p_\nu p^\prime_\mu) \nn \\
&&\qq +i\,
T_1^{B_c \to V}(q^2) (p_\mu \epsilon_\nu^*-\epsilon_\mu^* p_\nu) \nn \\
&&\qq +i\,T_2^{B_c \to V}(q^2)(p^\prime_\mu \epsilon_\nu^*-\epsilon_\mu^* p^\prime_\nu) , \nn
\eea
with the condition  
\bea  
A_0^{B_c \to V}(0)&=&  \frac{m_{B_c} + m_V}{2 m_V} A_1^{B_c \to V}(0) \nn \\
&-&  \frac{m_{B_c} - m_V}{2 m_V}  A_2^{B_c \to V}(0) . \qq \qq
\eea

The relations among the form factors and the universal functions  $\Omega_1(y)$ and $\Omega_2(y)$ are obtained using Eq.~\eqref{omega} \cite{Jenkins:1992nb}:
\bea
&&\langle P(v,k)| {\bar q} \gamma_\mu Q| { B_c}(v) \rangle = \nn \\
&&\qq  2 \sqrt{ m_{B_c} m_P} 
 \Big( \Omega_1(y) \ v_\mu + a_0 \Omega_2 (y) \ k_\mu \Big) , \nn \\
&&\langle P(v,k)| {\bar q} Q| { B_c}(v) \rangle =
2\sqrt{ m_{B_c}  m_P} \Big( \Omega_1(y) + a_0 \Omega_2(y) \ v \cdot k \Big), \nn  \\
&&\langle P(v,k)| {\bar q} \sigma_{\mu \nu } Q| {B_c}(v) \rangle =\nn \\
&&\qq  - 2 i  \sqrt{m_{B_c}  m_P} \ a_0 \Omega_2(y) \Big(v_\mu k_\nu - v_\nu k_\mu \Big),\label{BctoPHQ}
\eea
with $P=B_{s,d}$, $p=m_{B_c} v$, and $p^\prime=m_P v +k$, 
\bea
&&\langle V(v,k,\epsilon)|{\bar q} \gamma_\mu Q| {B_c}(v) \rangle = \nn \\
&& \,\,\,\,\,\,\,   2 i  \sqrt{ m_{B_c}  m_V} \ a_0 \Omega_2(y) \   \epsilon_{\mu \nu \alpha \beta} \epsilon^{*\nu} \ k^\alpha v^\beta, \nn \\
&&\langle V(v,k,\epsilon)|{\bar q} \gamma_\mu \gamma_5 b| { B_c}(v) \rangle = 2 \sqrt{ m_{B_c} m_V} \nn \\
&&\,\,\,\,\,\,\,  \Big(  \epsilon^{*}_{\mu} \, \left(\Omega_1(y) +v \cdot k  \ a_0 \Omega_2(y) \right) \nn \\ 
&&\,\,\,\,\,\,\,  -(v_\mu -\frac{k_\mu}{m_V} ) \epsilon^* \cdot k \,   a_0 \Omega_2(y)    \Big),  \nn \\ 
&&\langle V(v,k,\epsilon)|{\bar q} \sigma_{\mu \nu}  Q| {B_c}(v) \rangle =  - 2 \sqrt{ m_{B_c}  m_V} \nn \\
&& \Big(   \epsilon_{ \mu \nu \alpha \beta} \epsilon^{*\alpha} v^\beta \Omega_1(y) +  \epsilon_{ \mu \nu \alpha \beta} \epsilon^{*\alpha} k^\beta a_0 \Omega_2(y)  \Big),
\label{BctoVHQ} \qq \qq  \\
&&\langle V(v,k,\epsilon)|{\bar q} \sigma_{\mu \nu} \gamma_5  Q| {B_c}(v) \rangle = 
2 i \sqrt{ m_{B_c} m_V}  \nn \\
&&\,\,\,\,\,\,\,     \Big(  \epsilon^*_\nu ( v_\mu \Omega_1(y) + k_\mu a_0 \Omega_2(y))\nn \\
&&\,\,\,\,\,\,\,  - \epsilon^*_\mu ( v_\nu \Omega_1(y) + k_\nu a_0 \Omega_2(y) ) \Big) , \nn 
\eea
where $V=B^*_{s,d}$ and $y=1+ \frac{v \cdot k}{m_{P,V}}$. 
Invoking the HQ spin symmetry and comparing the first  equation in (\ref{BctoP}) to the corresponding one in (\ref{BctoPHQ}),
the form factors $\Omega_1$ and $\Omega_2$ are obtained   from $f_+$ and $f_0$:
\bea
\Omega_1&=&\frac{m_{B_c}+m_P}{2 q^2 \sqrt{m_{B_c} m_P}} \Big( (m_{B_c}-m_P)^2 (f_0-f_+) +q^2 f_+ \Big) \nn  \\
a_0 \Omega_2&=&\frac{1}{2 q^2 \sqrt{m_{B_c} m_P}} \Big( (m_{B_c}^2-m_P^2) (f_+-f_0) +q^2 f_+ \Big)
\nn \\ \label{eq:om12}
\eea
with $q^2=m_{B_c}^2+m_P^2-2 m_{B_c} m_P y$. These correspond to the results   in Fig.~\ref{fig:omega}.
Further comparing  (\ref{BctoP}) to  (\ref{BctoPHQ}), as well as (\ref{BctoV}) to (\ref{BctoVHQ}),  the  relations of all  form factors in terms of    $\Omega_{1,2}$ can be derived. For $B_c \to  P$ one has:
\bea
f_+^{B_c \to P}&=& \sqrt{\frac{m_P}{m_{B_c}} } \Big( \Omega_1+(m_{B_c}-m_P)  a_0 \Omega_2 \Big), \nn \\
f_0^{B_c \to P}&=&\sqrt{\frac{m_P}{m_{B_c}} } \frac{1}{m_{B_c}^2-m_P^2} \Big(  (m_{B_c}^2+q^2-m_P^2) \Omega_1 \nn \\
&+& (m_{B_c}+m_P)((m_{B_c}-m_P)^2  -q^2) a_0 \Omega_2  \Big),  \nn \\
f_T^{B_c \to P}&=&\sqrt{\frac{m_P}{m_{B_c}} }(m_{B_c}+m_P)  a_0 \Omega_2 \,\,. \label{A7}
\eea
%
For $B_c \to V$ one has:
\bea
&&V^{B_c \to V}=   \sqrt{\frac{m_V}{m_{B_c}} } (m_{B_c}+m_V)  a_0 \Omega_2 , \nn \\
&&A_0^{B_c \to V}= \nn \\
&&\qq \frac{1}{2 \sqrt{m_{B_c} m_V} } \Big( 2 m_{B_c} \Omega_1+(m_{B_c}^2-m_V^2+ q^2) a_0 \Omega_2\Big) , \nn \\
&&A_1^{B_c \to V}= 2 \sqrt{m_{B_c} m_V} \frac{1}{m_{B_c}+m_V} \Omega_1 , \nn \\
&&A_2^{B_c \to V}= - \sqrt{\frac{m_V}{m_{B_c}}} (m_{B_c}+m_V) a_0 \Omega_2  , \label{A8}\\
&&T_1^{B_c \to V}=2 \sqrt{\frac{m_V}{m_{B_c}}} \Big( \Omega_1  -m_V a_0 \Omega_2 \Big) , \nn \\
&&T_2^{B_c \to V}=2 \sqrt{m_{B_c} m_V} a_0 \Omega_2  ,\nn \\
&&T_0^{B_c \to V}=0 . \nn
\eea
Eqs.(\ref{A7})-(\ref{A8}) are obtained  for $v\cdot k =0$. Only $A_{0,1,2}$ are modified if this condition is not imposed,   the other relations remain unaffected.

\section{Coefficient functions in the  $B_c \to V(\to P\gamma) \, \bar\ell  \nu_\ell$ full angular distribution}\label{app:B}
In  Tables \ref{tab:SM}-\ref{tab:RPT} 
we collect the  functions $I_i$  in   Eq.~\eqref{angulargamma} for all  operators in the  Hamiltonian \eqref{hamil}, with $H_\pm, H_0, H_t$ and $H_\pm^{NP}, H_L^{NP}$  defined in Eqs.~\eqref{HampV}, \eqref{HampNP}.

 \begin{table}[h]
\caption{  \small  Angular coefficient functions  in the  decay distribution  Eq.~\eqref{angulargamma} for the Standard Model.}\label{tab:SM}
\vspace{0.3cm}
\centering
\begin{tabular}{cc}
\hline
\hline
\noalign{\medskip}
$i$ & $I_i^{\rm SM}$ \\
\noalign{\medskip}
\hline
\noalign{\smallskip}
$I_{1s}$ & $2 m_{\ell}^2 H_t^2 + H_0^2 (m_{\ell}^2 + q^2)$ \\
\noalign{\medskip}
$I_{1c}$ & $\frac{1}{8}(H_+^2 + H_-^2)(m_{\ell}^2 + 3 q^2)$ \\
\noalign{\medskip}
$I_{2s}$ & $ H_0^2 (m_{\ell}^2 - q^2)$ \\
\noalign{\medskip}
$I_{2c}$ & $- \frac{1}{8}(H_+^2 + H_-^2)(m_{\ell}^2 - q^2)$ \\
\noalign{\medskip}
$I_{3}$ & $ H_+ H_- (q^2-m_{\ell}^2)$  \\
\noalign{\medskip}
$I_{4}$ &$- \frac{1}{2} H_0 (H_+ + H_-) (m_{\ell}^2 - q^2)$\\
\noalign{\medskip}
$I_{5}$ & $ H_t (H_+ + H_-) m_{\ell}^2 +H_0 (H_+ - H_-) q^2$\\
\noalign{\medskip}
$I_{6s}$ & $- 4 H_t H_0 m_{\ell}^2$ \\
\noalign{\medskip}
$I_{6c}$ & $ \frac{1}{2} (H_+^2 - H_-^2) q^2$ \\
\noalign{\medskip}
$I_{7,8,9}$ & $0$ \\
\noalign{\medskip}
\hline
\hline
\end{tabular}
\end{table}

\begin{table}[h]
\caption{  \small Angular coefficient functions   in NP  with the operator ${\cal O}_R$ and   interference  SM-R  terms. The functions $I_i^{\rm R}$ are obtained from the corresponding SM functions replacing $H_+ \leftrightarrow H_-$.}\label{tab:R}
\vspace{0.3cm}
\centering
\begin{tabular}{ccc}
\hline
\hline
\noalign{\medskip}
$i$ &  $I_i^{\rm R}$& $I_i^{\rm INT,R}$ \\
\noalign{\medskip}
\hline
\noalign{\smallskip}
$I_{1s}$  &$2 m_{\ell}^2 H_t^2 + H_0^2 (m_{\ell}^2 + q^2)$&$-2 m_{\ell}^2 H_t^2 - H_0^2 (m_{\ell}^2 + q^2)$\\
\noalign{\medskip}
$I_{1c}$  & $\frac{1}{8}(H_+^2 + H_-^2)(m_{\ell}^2 + 3 q^2)$& $-\frac{1}{4}H_+ H_-(m_{\ell}^2 + 3 q^2)$\\
\noalign{\medskip}
$I_{2s}$  & $ H_0^2 (m_{\ell}^2 - q^2)$& $ -H_0^2 (m_{\ell}^2 - q^2)$\\
\noalign{\medskip}
$I_{2c}$ & $- \frac{1}{8}(H_+^2 + H_-^2)(m_{\ell}^2 - q^2)$& $ \frac{1}{4}H_+ H_-(m_{\ell}^2 - q^2)$ \\
\noalign{\medskip}
$I_{3}$ & $ H_+ H_- (q^2-m_{\ell}^2)$& $\frac{1}{2}(H_+^2 + H_-^2)(m_{\ell}^2 -q^2)$ \\
\noalign{\medskip}
$I_{4}$ &$- \frac{1}{2} H_0 (H_+ + H_-) (m_{\ell}^2 - q^2)$ &$ \frac{1}{2} H_0 (H_+ + H_-) (m_{\ell}^2 - q^2)$\\
\noalign{\medskip}
$I_{5}$ & $ H_t (H_+ + H_-) m_{\ell}^2 $& $ -H_t (H_+ + H_-) m_{\ell}^2 $\\
& $-H_0 (H_+ - H_-) q^2$ & \\
\noalign{\medskip}
$I_{6s}$  & $- 4 H_t H_0 m_{\ell}^2$& $ 4 H_t H_0 m_{\ell}^2$\\
\noalign{\medskip}
$I_{6c}$ & $ - \frac{1}{2} (H_+^2 - H_-^2) q^2$ & 0 \\
\noalign{\medskip}
$I_{7}$ & $0$& $ -H_t (H_+ - H_-) m_{\ell}^2 $ \\
\noalign{\medskip}
$I_{8}$ & $0$& $ \frac{1}{2} H_0 (H_+ - H_-) (m_{\ell}^2 - q^2)$\\
\noalign{\medskip}
$I_{9}$ & $0$& $\frac{1}{2}(H_+^2 - H_-^2)(m_{\ell}^2 -q^2)$ \\
\noalign{\medskip}
\hline
\hline
\end{tabular}
\end{table}
\begin{table}[htp]
\caption{\small Angular coefficient functions  for  NP  with the pseudoscalar P operator,  and  interference  SM-P terms.}\label{tab:P}
\vspace{0.3cm}
\centering
\begin{tabular}{ccc}
\hline
\hline
\noalign{\medskip}
$i$ & $I_i^{\np,P}$ & $I_i^{\inter,P}$  \\
\noalign{\medskip}
\hline
\noalign{\medskip}
$I_{1s}$ &$2 H_t^2 \frac{q^4}{(m_Q + m_q)^2}$ & $2 H_t^2 \frac{m_{\ell} q^2}{m_Q + m_q}$  \\
\noalign{\medskip}
$I_{1c,2s,2c,6c,3,4,8,9}$ & $0$ & $0$   \\
\noalign{\medskip}
$I_{5}$ & $0$ & $H_t(H_+ + H_-) \frac{m_{\ell} q^2}{2 (m_Q + m_q)}$    \\
\noalign{\medskip}
$I_{6s}$ & $0$ & $-2 H_t H_0 \frac{m_{\ell} q^2}{m_Q + m_q}$ \\
\noalign{\medskip}
$I_{7}$ & $0$ & $ H_t (H_+ - H_-) \frac{m_{\ell} q^2}{2(m_Q + m_q)}$  \\
\noalign{\smallskip}
\hline
\hline
\end{tabular}
\end{table}

\begin{table*}[t]
\caption{ \small Angular coefficient functions for NP with the tensor T operator   and  interference  SM-T terms.}\label{tab:T}
\vspace{0.3cm}
\centering
\begin{tabular}{ccc}
\hline
\hline
\noalign{\medskip}
$i$ & $I_i^{\rm NP,T}$ & $I_i^{\rm INT, T}$ \\
\noalign{\medskip}
\hline
\noalign{\bigskip}
$I_{1s}$ & $\frac{1}{16} (H_L^\np)^2 (m_\ell^2 + q^2)$ & $- \frac{1}{2}H_L^\np H_0 m_\ell \sqrt{q^2}$ \\
\noalign{\medskip}
$I_{1c}$ & $\frac{1}{2}[(H_+^\np)^2+(H_-^\np)^2](3 m_\ell^2 + q^2)$ & $- ( H_+^\np H_+ + H_-^\np H_- ) m_\ell \sqrt{q^2}$ \\
\noalign{\medskip}
$I_{2s}$ & $\frac{1}{16} (H_L^\np)^2 (q^2 - m_\ell^2)$ & $0$ \\
\noalign{\medskip}
$I_{2c}$ & $\frac{1}{2}[(H_+^\np)^2+(H_-^\np)^2](m_\ell^2 - q^2)$ & $0$ \\
\noalign{\medskip}
$I_{3}$ & $-4 H_+^\np H_-^\np (q^2 - m_\ell^2)$ & $0$ \\
\noalign{\medskip}
$I_{4}$ & $-\frac{1}{4} H_L^\np (H_+^\np + H_-^\np) (q^2 - m_\ell^2)$ & $0$ \\
\noalign{\medskip}
$I_{5}$ &$\frac{1}{2}  H_L^\np (H_+^\np - H_-^\np) m_\ell^2$ & $-\frac{1}{8} [H_L^\np (H_+ - H_-) + 8 H_+^\np (H_t + H_0)  + 8 H_-^\np(H_t - H_0)] m_{\ell} \sqrt{q^2}$ \\
\noalign{\medskip}
$I_{6s}$ & $0$ &  $\frac{1}{2}H_L^\np H_t m_{\ell} \sqrt{q^2}$\\
\noalign{\medskip}
$I_{6c}$ & $2[(H_+^\np)^2-(H_-^\np)^2]m_\ell^2$& $- ( H_+^\np H_+ - H_-^\np H_- ) m_\ell \sqrt{q^2}$ \\
\noalign{\medskip}
$I_{7}$ & $0$ & $-\frac{1}{8} [H_L^\np (H_+ + H_-) - 8 H_+^\np (H_t + H_0)  + 8 H_-^\np(H_t - H_0)] m_{\ell} \sqrt{q^2}$ \\
\noalign{\medskip}
$I_{8,9}$ & $0$ & $0$ \\
\noalign{\smallskip}
\hline
\hline
\end{tabular}
\end{table*}

\begin{table*}[t]
\caption{\small P-R, R-T and P-T interference terms in the angular coefficient functions. }\label{tab:RPT}
\vspace{0.3cm}
\centering
\begin{tabular}{cccc}
\hline
\hline
\noalign{\medskip}
$i$ & $I_i^{\inter,PR}$ & $I_i^{\inter,RT}$ & $I_i^{INT, PT}$ \\
\noalign{\medskip}
\hline
\noalign{\medskip}
$I_{1s}$ & $-2 H_t^2 \frac{m_\ell q^2}{(m_b + m_q)}$ & $\frac{1}{2}H_0 H_L^\np m_\ell \sqrt{q^2}$ & $0$\\
\noalign{\medskip}
$I_{1c}$ & $0$ & $(H_+^{NP}H_- + H_-^{NP}H_+) m_{\ell} \sqrt{q^2}$ & $0$\\
\noalign{\medskip}
$I_{2s,2c,3,4,8,9}$ & $0$ & $0$ & $0$ \\
\noalign{\medskip}
$I_{5}$ & $-H_t(H_+ + H_-) \frac{m_{\ell} q^2}{2(m_Q + m_q)}$ & $\frac{1}{8} [H_L^\np (H_- - H_+) + 8 H_+^\np (H_t + H_0)  $ &\,\,  $ -H_t(H_+^{NP} + H_-^{NP}) \frac{ (q^2)^{3/2}}{m_Q + m_q}$   \\
\noalign{\medskip}
$\quad$ & $\quad$ & $\qquad \qquad + 8 H_-^\np(H_t - H_0)] m_{\ell} \sqrt{q^2}$ & $\quad$\\
\noalign{\medskip}
$I_{6s}$ & $2 H_t H_0 \frac{m_{\ell} q^2}{m_Q + m_q}$ & $-\frac{1}{2}H_t\, H_L^{NP} m_\ell \sqrt{q^2}$ &  \,\,$ H_t\, H_L^{NP} \frac{ (q^2)^{3/2}}{2(m_Q + m_q)}$\\
\noalign{\medskip}
$I_{6c}$ & $0$ & $(H_+^{NP}H_- - H_-^{NP}H_+) m_{\ell} \sqrt{q^2}$ &$0$\\
\noalign{\medskip}
$I_{7}$ & $ H_t (H_+ - H_-) \frac{m_{\ell} q^2}{2(m_Q + m_q)}$& $-\frac{1}{8} [H_L^\np (H_- + H_+) - 8 H_+^\np (H_t + H_0)  $  & \,\, $ -H_t(H_+^{NP} - H_-^{NP}) \frac{ (q^2)^{3/2}}{m_Q + m_q}$ \\
\noalign{\medskip}
$\quad$ & $\quad$ & $\qquad \qquad + 8 H_-^\np(H_t - H_0)] m_{\ell} \sqrt{q^2}$ & $\quad$\\
\noalign{\smallskip}
\hline
\hline
\end{tabular}
\end{table*}

\newpage
\bibliographystyle{apsrev4-1}
\bibliography{refsFFP4}
\end{document}